\documentclass[manuscript, screen, nonacm]{acmart}

\usepackage{multirow}
\usepackage{booktabs}
\usepackage{enumitem}
\usepackage{graphicx}
\usepackage{url}
\usepackage{textcomp}
\usepackage{xcolor}
\usepackage{listings}
\usepackage[most]{tcolorbox}
\usepackage{makecell}
\usepackage{longtable}

\AtBeginDocument{%
  }

\setcopyright{none}
\acmDOI{}
\acmISBN{}

\begin{document}

\title{How Safe Are AI-Generated Patches? A Large-scale Study on Security Risks in LLM and Agentic Automated Program Repair on SWE-bench}

\author{Amirali Sajadi}
\email{amirali.sajadi@drexel.edu}
\orcid{https://orcid.org/0000-0003-0188-7005}
\affiliation{%
  \institution{Drexel University}
  \city{Philadelphia}
  \state{Pennsylvania}
  \country{USA}
}

\author{Kostadin Damevski}
\email{kdamevski@vcu.edu}
\orcid{https://orcid.org/0000-0001-7799-2026}
\affiliation{%
  \institution{Virginia Commonwealth University}
  \city{Richmond}
  \state{Virginia}
  \country{USA}
}

\author{Preetha Chatterjee}
\email{preetha.chatterjee@drexel.edu}
\orcid{https://orcid.org/0000-0003-3057-7807}
\affiliation{%
  \institution{Drexel University}
  \city{Philadelphia}
  \state{Pennsylvania}
  \country{USA}
}

\renewcommand{\shortauthors}{Sajadi et al.}

\begin{abstract}
Large language models (LLMs) and agentic frameworks are increasingly used for development tasks such as automated program repair (APR). Prior work has identified security risks in LLM-generated code, but most studies focus on synthetic, simplified, or isolated tasks. We present a large-scale security analysis of LLM-generated patches using 20,000+ GitHub issues from SWE-bench. We evaluate developer-written patches, two standalone LLMs (Llama 3.3 Instruct-70B and GPT-5.4 nano), and four agentic frameworks (OpenHands, AutoCodeRover, HoneyComb, and SWE-agent). We also analyze code, issue, and project-level factors associated with insecure patches.

Under our evaluated SWE-bench setup, Llama 3.3 introduced 135 manually confirmed vulnerabilities across the train and test sets, compared with 12 in developer-written patches for the same issues. On the SWE-bench test set, GPT-5.4 nano introduced 3 vulnerabilities compared with 2 in developer patches. All four evaluated agentic frameworks also introduced vulnerabilities. OpenHands had the highest retained count, with 44 vulnerabilities across 11 issue instances. Vulnerabilities are associated with broader code modifications and with issues missing specific information. These results suggest that APR systems require proactive risk assessment methods that consider both generated code and issue context.
\end{abstract}


\maketitle

\section{Introduction}

LLMs are rapidly transforming software development, assisting with tasks like code completion, test generation, and bug fixing~\cite{li2023cctest, siddiq2024using, zhang2023well, alshahwan2024automated, wei2023copiloting}. They are also increasingly integrated into agentic frameworks that enable autonomous planning and execution of complex tasks, such as automated program repair (APR), where LLM-based agents identify, diagnose, and patch software bugs directly in real-world projects~\cite{openhands, yang2024sweagent, zhang2024autocoderover, 10.1145/3832783.3837434}. These systems have significantly outperformed traditional APR approaches and are now used to solve repairs across numerous real-world software projects, spanning thousands of issues from production-grade libraries and frameworks~\cite{11334717}.
This widespread adoption of LLMs in APR systems raises critical concerns about their potential security implications \cite{perry2023users, yao2024survey, sajadi2025llms, siddiq2023generate}. Studies show that 30-50\% of code generated with LLM-based code assistants contains  vulnerabilities~\cite{pearce2025asleep, fu2023security, wang2024aigeneratedcodereallysafe}.
When embedded within automated program repair workflows, LLM-generated patches can be routed into testing, review, or CI/CD workflows, where an insecure repair can still propagate if it is accepted without sufficient security review.

While several studies have shown that LLMs are prone to generating insecure or low-quality code, most focus on simplified tasks such as coding exercises or identifying vulnerabilities in isolated code snippets~\cite{mohsin2024can, siddiq2024quality, pearce2025asleep, khoury2023secure}. These scenarios lack the complexity and context of program repair in real projects, where understanding environment constraints or project-specific dependencies is critical. Moreover, such scenarios use LLMs as generic code generators rather than APR systems that produce targeted edits for specific bugs. APR agents also operate in a different setting than ordinary code agents. Code AI agents generate code from a short user prompt, while APR agents take a bug report, inspect the current repository state, and then propose candidate patches to fix a specific bug~\cite{openhands, yang2024sweagent, zhang2024autocoderover}. Hence, assessing their security requires evaluation frameworks that capture the full end-to-end repair workflow rather than isolated snippet-based analyses~\cite{sajadi2026traceviewinteractivevisualizationagentic}.

Although recent benchmarks like SWE-bench \cite{jimenez2024swebench} have enabled the assessment of LLMs on real-world software engineering tasks, security-specific analyses in these contexts remain limited, and we still lack a systematic understanding of when APR-style LLM and agentic frameworks introduce new vulnerabilities, or how those vulnerabilities relate to properties of the patches and the surrounding development context. Recent studies have demonstrated that adversarial bug reports can mislead APR agents into producing insecure yet functionally correct patches~\cite{przymus2025adversarialbugreportssecurity, chen2025redteamingprogramrepair}. These efforts evaluate security risks only assuming adversarial intent and on relatively small datasets. No prior work has performed a large-scale and security-focused empirical analysis of patches generated by LLM and agentic APR systems, especially assessing if vulnerabilities in patches arise even without adversarial intent.

Using SWE-bench, we analyze patches generated by Llama 3.3 and GPT-5.4 nano and four agentic APR frameworks, OpenHands, AutoCodeRover, HoneyComb, and SWE-agent, as well as the developer-written fixes for the same issues. We first quantify how often these systems introduce new vulnerabilities and which CWE types they tend to produce. We then examine how code-level, issue-level, and project-level factors are associated with the security of the proposed solutions, identifying contexts in which vulnerability-inducing AI-generated patches occur. Our overarching goal with this work is to identify when and where APR systems are most likely to generate vulnerable patches, informing guidelines for when to trust their output and when to require additional checks. Overall, these analyses provide a large-scale empirical evidence of the conditions and the frequency with which APR systems introduce security vulnerabilities when resolving real GitHub issues. Our contribution is therefore not the general observation that LLMs can generate insecure code, but the evaluation of this risk in realistic APR workflows and the identification of recurring contextual patterns that can help developers and future APR systems avoid insecure generated patches. We answer the following research questions:

\begin{itemize}
    \item \textbf{RQ1}: How secure are the patches produced by \textbf{standalone LLMs} on real-world issue resolution tasks? How does their security compare to developer-written patches?
    \item \textbf{RQ2}: How secure are the patches produced by \textbf{agentic APR workflows} on real-world issue resolution tasks? How does their security compare to developer-written patches?
    \item \textbf{RQ3}: What code, issue, or project characteristics are associated with the generation of vulnerable code by LLMs or agentic APR frameworks when resolving issues?
\end{itemize}
Our findings highlight the security risks of LLM-generated code at scale. In both evaluated standalone models, we find manually confirmed vulnerabilities. Under our SWE-bench train+test evaluation setup covering 20,000+ GitHub issues, Llama 3.3 introduced 135 new vulnerabilities, approximately 11 times the 12 new vulnerabilities introduced by developers on the same issues. GPT-5.4 nano introduced manually confirmed vulnerabilities on the SWE-bench test set. Across the SWE-bench test set, all four evaluated agentic frameworks produced manually confirmed vulnerabilities. OpenHands had the highest retained count, with 35 of its 44 vulnerabilities concentrated in three issues. Notably, Llama 3.3 produced more retained findings in categories such as CWE-95 and CWE-502 than developers in this setup; the corresponding developer patches also included categories such as CWE-732 and CWE-377. We found that vulnerable outputs are often associated with broader code modifications, especially changes that span more files. Further, most retained Llama 3.3 vulnerabilities were observed in instances involving bug-related issues that do not contain code examples, information about the expected behavior of the code, or steps to reproduce the issue. These findings suggest that the risk of generating vulnerabilities is not only a function of the code produced but also of the context in which the LLM operates. As such, our insights can help inform proactive risk assessments of LLM and agent contributions, complementing the existing vulnerability detection tools, by \textit{highlighting high-risk repair tasks even before the code is generated or reviewed.}

\section{Methodology}
\label{methodology}
\noindent
Figure \ref{fig:method-main} shows an overview of our methodology. Using GitHub issues from SWE-bench, we collect developer patches and candidate patches generated by two standalone LLMs (Llama 3.3~\cite{meta2025llama3} and GPT-5.4 nano~\cite{openai_gpt54nano}) and four agentic frameworks (OpenHands~\cite{openhands}, AutoCodeRover~\cite{zhang2024autocoderover}, HoneyComb~\cite{honeycomb}, and SWE-agent~\cite{yang2024sweagent}). We analyze the prevalence and patterns of vulnerabilities for each of these code generation sources (RQ1, RQ2). In addition, we identify code, issue, and project-level factors that are associated with the presence of these vulnerabilities (RQ3). Our full data and research artifacts are publicly accessible through our replication package~\cite{llm_security2025}\footnote{a lightweight repository containing the analysis code, scripts, and documentation without the large generated datasets is available at \url{https://anonymous.4open.science/r/SSWE_Replication_Data_lightweight-2A65}.}.

\subsection{Dataset and Task Setup}
\label{dataset}
We leverage the SWE-bench dataset \cite{jimenez2024swebench}, which consists of real-world GitHub issues paired with pull requests that resolve them. SWE-bench includes a \textit{train set} of approximately 19,000 issue--PR pairs collected from 37 popular Python repositories and a \textit{test set} of 2,294 issue--PR pairs from 12 repositories.

For the standalone LLMs, we use Llama 3.3 Instruct (70B) to generate patches for both the train and test sets of SWE-bench, comprising approximately 21,000 instances in total. Since Llama is open-source and possible to run at scale, this generation step, although computationally heavy, is feasible. {We use GPT-5.4 nano~\cite{openai_gpt54nano} on the SWE-bench test set as a recent small commercial standalone LLM. We selected GPT-5.4 nano because it was the newest small API-based GPT model available for our study, while still being financially feasible for test-set patch generation. For the agentic frameworks, we rely on publicly released patches that are available only for the test set~\cite{swebench2023experiments}. Due to the substantial computational and financial cost required to run these iterative frameworks across the train set, we restrict our framework-based analysis to the test split.

Including both standalone LLM and agentic framework settings ensures the correct scope for addressing our research questions (RQ1--RQ2). These two settings also reflect distinct real-world usage scenarios. Standalone LLMs can better represent how developers interact with models, prompting them to get help with issue resolution and manually integrating the generated code. This makes the RQ1 analysis relevant for understanding the security implications of everyday developer usage. Agentic frameworks, in contrast, embody more autonomous approaches to issue resolution and reflect a growing trend toward automated, end-to-end program repair. Evaluating both settings enables a broader understanding of the security risks posed by current and emerging LLM-based workflows. Our evaluation analyzes the patches produced by the models and agents before any human review or refinement. In practice, developers may inspect and revise these patches before merging them, but including such revisions would conflate the security of the tool output with the effects of human review. We therefore measure the baseline security behavior of the models and agents as standalone repair systems.

\begin{figure*}[t]
    \centering
    \includegraphics[width=\textwidth]{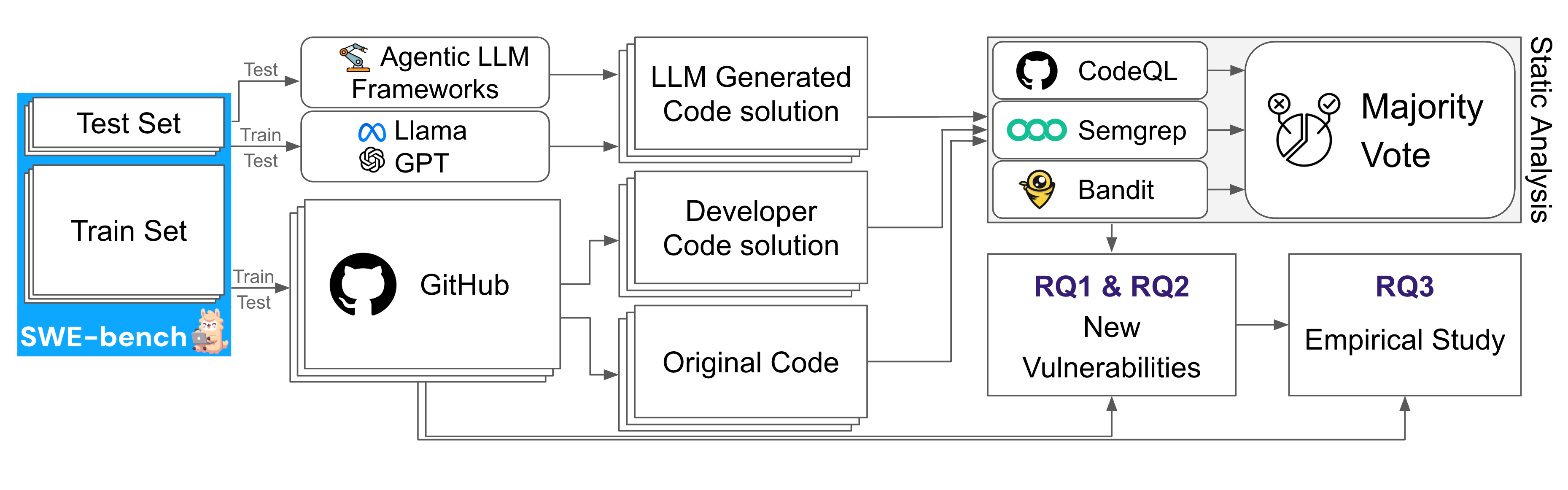}
    \vspace{-2em}
    \caption{Overview of Our Methodology for LLM-Based Code Generation and Security Evaluation.}
    \label{fig:method-main}
    \vspace{-.6cm}
\end{figure*}

\subsection{Patch Generation with LLM and Agentic Frameworks}  

\subsubsection{\textbf{RQ1.} Standalone LLMs}
To examine the security risks of patches produced by standalone LLMs, we generate candidate patches across the entire dataset (train+test) using the Llama 3.3 Instruct 70B model \cite{meta2025llama3}. Recent benchmarks indicate that this model's code generation capabilities are comparable to those of GPT-4o \cite{openai2025gpt4o} and exceed the performance of Claude 3 Opus \cite{anthropic_claude3_family} and GPT-4 Turbo \cite{openai2023gpt4turbo}, making it a strong choice for our full-scale study~\cite{lmarena2024leaderboard, chiang2024chatbot}. We selected Llama 3.3 instead of the Code Llama family because recent general-purpose Llama models report stronger coding benchmark performance than Code Llama~\cite{llama33modelcard, roziere2023code}. Further, due to the computational cost of generating patches for over 20,000 issues, evaluating additional 70B-scale models on the full dataset is not feasible. We also evaluate GPT-5.4 nano on the SWE-bench test split, allowing us to evaluate a recent commercial standalone LLM.

\textit{Bug Localization:}
We use the oracle retrieval method, i.e., supplying the model with the corresponding issue description and the exact project files modified in the developer's ground-truth fix. This ensures that the model receives the most relevant parts of the codebase that are needed to address the issue. We intentionally adopt the oracle retrieval method since it allows us to isolate and evaluate the security of the code generated by the LLM itself, without conflating generation quality with the efficacy of retrieval methods. Additionally, the default SWE-bench retrieval baseline, BM25, fails to retrieve any relevant files for approximately 40\% of test instances~\cite{jimenez2024swebench}. The focus of our study is on assessing the security of generated code when an LLM is provided with the correct context.

\textit{LLM Prompts:}
We adopt the prompt structure used in SWE-bench, with slight modifications. Rather than generating diffs like SWE-bench, we explicitly instruct the model to produce complete files. While diff generation is popular for its lower token overhead and ease of use with patching tools~\cite{wang2025solved}, in standalone LLM settings, it often yields syntactically invalid or partially applicable patches, even after post-processing~\cite{jimenez2024swebench}. 
Our preliminary experiments confirmed this limitation. In contrast, producing full files ensures self-contained outputs without relying on diff syntax. It also reflects realistic usage, where developers expect standalone models to produce runnable code, not diffs. For this study, we prioritize reliability over minimal token usage. 

\subsubsection{\textbf{RQ2.} Agentic LLM Frameworks}
To investigate whether agentic workflows introduce security risks in LLM-generated patches, we analyze solutions from four high-performing agentic APR framework outputs evaluated on the SWE-bench Full test set: \textit{OpenHands+CodeAct v2.1 (Claude-3.5-Sonnet-20241022)} \cite{openhands}, \textit{AutoCodeRover-v2.0 (Claude-3.5-Sonnet-20241022)} \cite{zhang2024autocoderover}, \textit{HoneyComb} \cite{honeycomb}, and \textit{SWE-agent 1.0 (Claude 3.7 Sonnet)}~\cite{yang2024sweagent, anthropic_claude3_family}. We select these systems because they provide publicly available outputs with strong functional performance on the same SWE-bench Full test split. These framework-model combinations resolved 29.38\%, 24.89\%, 22.06\%, and 33.83\% of the SWE-bench Full test set issues, respectively. We therefore treat these systems as a relevant, but not exhaustive, set of agentic APR outputs under the same benchmark setting.

To enable a uniform comparison across models, we modified and extended the official SWE-bench evaluation harness. This harness typically applies a candidate patch to a Dockerized snapshot of the original repository state and runs corresponding test cases to assess functional correctness. For our purposes, we modified this process to capture the full impact of each patch on the project's files. Specifically, for every file modified by a patch, we extract and save two versions: one pre-patch snapshot and one post-patch snapshot. We also track newly added and renamed files with their full file paths. This is necessary because a patch may rename a file, causing the pre-patch and post-patch versions of the same file to appear under different paths. Tracking these paths allows us to map the before and after versions consistently during later vulnerability comparisons. This process yields two datasets per framework: one consisting of pre-patch files and another consisting of post-patch files. These datasets capture the complete file-level changes made by the framework while applying the patches.

It is important to emphasize that our RQ2 evaluation is not a direct comparison between agentic frameworks and standalone LLMs. OpenHands, AutoCodeRover, HoneyComb, and SWE-agent are evaluated using the test split, whereas the Llama and developer patches are analyzed on the combined train+test set. GPT-5.4 nano is also evaluated on the test split. Our research questions are intentionally scoped to characterize behavior within each setting. For standalone LLMs, we measure how often their patches introduce vulnerabilities compared to developers on the same issues; for the agentic frameworks, we ask whether they also introduce vulnerabilities on the subset of issues they attempt to resolve, and how their observed vulnerabilities compare descriptively to developer-written patches on the same test-set tasks. We therefore restrict all quantitative conclusions to the datasets on which each method is actually evaluated and avoid any claims of statistical superiority or inferiority between the standalone LLMs, the agentic frameworks, and developer-written patches in the test-set comparisons.

\subsection{Developer-written Patch Collection}
To create a security baseline for comparison with LLM-generated code, we collect the original developer-written solutions for each issue from the corresponding PR using the GitHub API \cite{github_api}. For each instance, we retrieve the final version of the files modified or created by the developer as part of the PR and merged into the codebase.

\subsection{Vulnerability Detection}
Manually analyzing the security of 20,000+ patches across multiple sources is infeasible. Therefore, in line with prior studies~\cite{siddiq2024sallm, hamer2024just, siddiq2024quality, della2025prompt}, we utilize established OWASP-recommended~\cite{owasp_source_code_analysis} static analysis tools to automate the vulnerability detection process. The following tools were selected based on their widespread use~\cite{siddiq2024quality, kree2024using, siddiq2023generate}, support for both rule-based and taint analysis, and their capacity to detect a wide range of real-world vulnerabilities \cite{li2025iris}:

\textbf{CodeQL}~\cite{github_codeql} is a static analysis tool developed by GitHub that is often used to identify security issues and code quality problems through heuristic and taint analysis techniques.

\textbf{Semgrep}~\cite{semgrep} is a static analysis tool that uses pattern-based rules and taint analysis to detect a wide range of vulnerabilities, e.g., injection risks and insecure API usage.

\textbf{Bandit}\cite{bandit} is a heuristic-based and security-focused static analysis tool designed for Python projects. Bandit inspects abstract syntax trees to detect common security issues.

When analyzing patches, we run each static analysis tool on both the original (pre-patch) and modified (post-patch) files. By comparing the two, we classify findings as either \textit{persisting vulnerabilities} (present in both the original and modified code) or \textit{new vulnerabilities} (introduced only after the patch). Only the latter are counted in our results, as they directly capture the security risks added as a result of the modifications.

A common concern with static analysis tools is their accuracy, particularly their tendency to produce false positives \cite{medeiros2014automatic, guo2023mitigating, dimastrogiovanni2016towards, nadeem2012high}. To overcome any shortcomings in the recall and precision of static analysis tools, we create a multi-step process as follows:

\textbf{Step 1: To maximize coverage (recall),} we combined the results of three static analysis tools, in line with prior studies~\cite{shahid2025llm}, since it has been shown that different tools have complementary strengths and that combining them provides a more comprehensive assessment of code quality~\cite{nunes2017combining, trautsch2023automated, charoenwet2024empirical, nunes2015phpsafe}. The findings of the static analysis tools are aggregated using their reported vulnerability types (CWEs). Using multiple tools in this way improves coverage and recall in two ways: (A) it broadens the set of CWE types we can observe, since any single tool is limited to the classes of vulnerabilities it supports, and (B) even when all three tools nominally cover the same CWE, any one of them can miss a concrete instance that another flags due to differences in internal analyses. All three tools perform pattern-based static analysis, but CodeQL and Semgrep additionally support taint analysis (relying on out-of-the-box taint specifications for popular APIs), and CodeQL can perform partial constraint reasoning. These complementary analyses mean that aggregating results across tools yields the highest number of true positive findings overall, thereby maximizing recall. Here, the file-level focus of the analyses not only distinguishes between new and pre-existing issues but also helps reconcile the fact that static analysis tools can report inconsistent information (e.g., severity levels, line ranges, or explanations) for the same underlying vulnerabilities~\cite{lenarduzzi2023critical}.

\textbf{Step 2: To increase confidence (precision),} we adopt a conservative \textit{majority voting strategy}: a generated or developer-written patch is treated as a security risk only if at least two of the three tools independently identify new vulnerabilities in the same file(s) it modifies or creates. Moving beyond aggregation, the majority vote dismisses findings reported by only a single tool as likely false positives. Prior work has shown that a higher concentration of static analysis warnings correlates with an increased likelihood of risky code, and that considering such concentrations can improve precision in practice~\cite{trautsch2023automated, charoenwet2024empirical}. Additionally, merging results from multiple static analysis tools has been shown to significantly reduce the likelihood of incorrect detections~\cite{dyremark2025false, meng2008approach, algaith2018finding}. Accordingly, independent agreement across tools both strengthens the case that a patch should be treated with caution and reduces false positives, thereby improving precision while retaining the broader coverage benefits of combining multiple tools. This conservative approach can provide a balanced and trustworthy basis for our large-scale comparative analysis across different patch sources, including developer patches, standalone LLMs, and agentic frameworks.

File-level selection accommodates differences in where tools report related security behavior. In the code excerpt in Figure~\ref{fig:serialization-warnings}, Semgrep flags serialization through \texttt{marshal.dumps()} at line 66, while Bandit flags deserialization of the same return value through \texttt{marshal.loads()} at line 97. These reports concern different stages of one data flow. Requiring reports to identify the same line or function could exclude such related candidates, whereas file-level selection accommodates these reporting differences.

\begin{figure}[ht]
\centering
\begin{minipage}{\linewidth}
\renewcommand{\thelstnumber}{%
    \ifnum\value{lstnumber}=67\relax
    \else\arabic{lstnumber}\fi}
\begin{lstlisting}[
    language=Python,
    numbers=left,
    firstnumber=66,
    numberstyle=\scriptsize\ttfamily,
    numbersep=6pt,
    escapeinside={(*@}{@*)}
]
retvalf.write(marshal.dumps(retval))  (*@\textcolor{red}{\# Semgrep: serialize return value}@*)
...(*@\setcounter{lstnumber}{96}@*)
retval = marshal.loads(retval_data)  (*@\textcolor{red}{\# Bandit: deserialize return value}@*)
\end{lstlisting}
\end{minipage}
\caption{Reports at different stages of the same data flow in OpenHands's \texttt{pytest-dev\_\_pytest-5787} patch. The child serializes its return value to a file that the parent subsequently reads and deserializes. Original line numbers are preserved; omitted code is indicated by an ellipsis.}
\label{fig:serialization-warnings}
\end{figure}

\textbf{Step 3: To ensure validity of all instances} and to guarantee that only valid security-relevant issues enter our analysis, \textbf{we performed a manual inspection} in which the first author inspected the patched file, and judged whether the tool-reported issue constituted a plausible security threat rather than a benign code smell. Each finding in a majority-selected file was assessed individually, and false positives were discarded. A finding meeting our validation criteria was retained even if another tool's report in the same file was rejected. Throughout the manual inspection, whenever a finding occurred inside a function, we conservatively assumed attacker-controlled input could reach the function. The instance was retained only if no adequate input validation or sanitization was evident within the function, along the path to the relevant vulnerable lines (e.g., whitelists, length/range checks, special-character checks, or calls to dedicated sanitizers). We excluded cases where inputs were hard-coded to safe values in one-off scripts or otherwise not reusable as library functionality. We also excluded findings in test-only files, reproduction scripts, debug or scratch files, vendored dependencies, and one-off helper artifacts when they were not intended to become reusable project or library code. We also evaluated the security of instances in light of intent. For example, instances labeled for use of \texttt{SHA-1} were removed when hashing served non-cryptographic purposes (e.g., content addressing). Conversely, we retained findings when untrusted parameters could influence critical operations, such as command execution (\texttt{shell=True} with function arguments), deserialization, path construction, or SQL string creation, and no adequate sanitization was evident. 

This multi-step process addresses known limitations of fully automated static analysis and ensures the quality of our findings prior to our quantitative analyses. Our pipeline combines Bandit’s syntax-based checks with Semgrep’s pattern-based and taint analysis and CodeQL’s data-flow analysis. For example, CodeQL can identify command injection by tracking whether untrusted input reaches a shell-execution operation without appropriate sanitization. This relationship can be detected across different variable names, intermediate assignments, and supported function calls, regardless of whether the code was generated by an LLM or written by a developer. Thus, detection depends on the underlying security-relevant operations and data-flow relationships supported by the tools, rather than on how the code is written.

\subsection{Metrics for Empirical Analysis}
We analyze the prevalence and properties of vulnerabilities (RQ1 \& RQ2) using CWE categories. We collect various code, issue, and project-level metrics to investigate the factors that are associated with the generation of vulnerabilities (RQ3).

\subsubsection{RQ1 \& RQ2: Vulnerability Prevalence and Properties}
We analyze the prevalence and types of vulnerabilities (identified via majority voting) in code generated by Llama 3.3 Instruct (70B) and GPT-5.4 nano and agentic frameworks (OpenHands, AutoCodeRover, HoneyComb, SWE-agent), and compare them to vulnerabilities in developer-written patches from GitHub. Specifically, we examine:

\textbf{Vulnerability prevalence:} We measure the number of GitHub issues containing at least one new vulnerability.

\textbf{Vulnerability types and overlaps:}  We measure the distribution of vulnerability types, based on CWE categories. Further, we investigate the similarities and differences between the vulnerability patterns in LLM and developer-written code.

\subsubsection{RQ3: Code, Issue, Project Characteristics Associated with LLM or agent-generated Vulnerabilities}
We investigate whether the context in which the model is tasked with resolving an issue is associated with the security of the generated code. We analyze a range of metrics that have been shown in prior work to correlate with the riskiness of code contributions \cite{goyal2018identifying, alali2008s, rosen2015commit, gonzalez2021anomalicious}, as well as introduce new metrics. Overall, these factors reflect different dimensions of task complexity and clarity. For instance, we hypothesize that issues that clearly and extensively describe the problem can be easier for LLMs to address, while tasks involving large or complex code changes may introduce more room for errors~\cite{ehsani2026characteristics}. In addition, characteristics of the project, such as size, complexity, and developer activity, may be associated with the likelihood of producing vulnerable code \cite{gkortzis2021software}. We perform this analysis for all vulnerabilities introduced both by standalone LLMs and the agentic workflows, allowing us to identify shared and distinct risk factors across different generation techniques. 

We group these factors into three categories: \textit{code-related}, \textit{issue-related}, and \textit{project-level} characteristics:

\textbf{Code-related Characteristics:} We analyze various properties of the LLM-generated code, including:

\noindent\underline{\textit{\# Files Added or Modified:}} A larger number of files can indicate broader changes and a higher overall risk \cite{goyal2018identifying, alali2008s, rosen2015commit, gonzalez2021anomalicious}.

\noindent\underline{\textit{\# Unique File Types Modified:}} Modifying diverse file types may introduce different complexities to the issue resolution process \cite{gonzalez2021anomalicious, goyal2018identifying}.

\noindent\underline{\textit{\# Sensitive File Modifications:}} Inspired by prior work on anomaly detecting malicious commits \cite{gonzalez2021anomalicious}, we define a list of sensitive file types (e.g., \texttt{.conf}, \texttt{.yml}, \texttt{.json}) whose modification we examine as a potential vulnerability risk indicator.

\noindent\underline{\textit{LOC Generated by LLM:}} This measures how many lines were added or rewritten by the LLM. Generating more lines of code may suggest that the model is undertaking more complex tasks, handling broader context, or replacing larger code chunks. We examine generation extent as a potential contextual risk indicator.

\noindent\underline{\textit{Cyclomatic Complexity of LLM-Generated Code:}} This metric reflects the complexity of the functions introduced or rewritten by the LLM. Prior studies also analyzed the cyclomatic complexity of LLM-generated code~\cite{fu2023security, toth2024llms}. Higher complexity in generated code indicates more intricate control flow. We examine complexity as a potential contextual risk indicator.

\textbf{Issue-related Characteristics}: We extract several features from the GitHub issue associated with each instance:

\noindent\underline{\textit{Issue Type and Bug Type Classification:}} We manually label whether the issue is a bug, a feature request, or a question, following established comprehensive taxonomies \cite{kallis2019ticket, song2020bee}. Further, for issues categorized as bugs, we classify the bug type using a comprehensive, empirically grounded, and widely-adopted taxonomy\cite{catolino2019not}. The bug categories are: \textit{Configuration Issues}, \textit{Network issues}, \textit{Database-related issues}, \textit{GUI-related issues}, \textit{Performance issues}, \textit{Permission/deprecation issues}, \textit{Security issues}, \textit{Program anomaly issues}, \textit{Test code-related issues}, and \textit{Miscellaneous/other issues}.

\noindent\underline{\textit{Information Completeness:}} Based on prior work emphasizing the importance of issue clarity \cite{song2020bee}, we manually annotate whether each issue includes: (1) \textit{Expected Behavior}: how the software is supposed to behave, (2) \textit{Observed Behavior}: how it actually behaves, and (3) \textit{Steps to Reproduce}: the process needed to trigger the issue. The presence of this information can aid in understanding the bug's context and identifying the root cause of the problem.

\noindent\underline{\textit{\# Words:}} We measure the total word count of the issue description and all the comments made on the issue prior to the date of first commit in the corresponding PR.

\noindent\underline{\textit{\# Comments Prior to First PR:}} We measure the number of comments on the issue thread made before the first pull request was submitted to address the problem. A higher comment count may indicate increased back-and-forth discussion, potentially reflecting communication challenges or ambiguities that required clarification through developer interactions.

\noindent\underline{\textit{Presence of Code Snippets:}} Using regular expressions, we identify code snippets in issues and use them as indicator of technical detail and potential guidance for bug resolution. 

Manual annotation of \textit{issue type}, \textit{bug type}, and \textit{information completeness} was performed independently by two annotators, each with at least three years of programming experience. For each data point, they examine both textual and code-related data, including the GitHub issue title, description, discussion thread, etc. The annotators followed a shared set of annotation instructions (included in our replication package). They followed an iterative analysis comprising multiple sessions. Inter-rater agreement was initially measured using Cohen's kappa for each category (all values $>$ 0.61). The annotation process proceeded through several rounds of labeling and discussion. This resulted in a final strong agreement level (Cohen's kappa values $>$ 0.9).

\textbf{Project-related Characteristics: }
Finally, we assess project characteristics associated with the vulnerable LLM patches:

\noindent\underline{\textit{\# Contributors:}} Using the GitHub API, we measure the number of developers contributing to a project. Specifically, we count all users listed as GitHub contributors, i.e., users who have made at least one commit to the repository's default branch. This metric reflects the breadth of developer involvement in a project. We examine whether contributor count is associated with vulnerability introduction. 

\noindent\underline{\textit{\# Files:}} The number of files in a given project can serve as a proxy for its potential complexity and size. Prior work on open-source software effort estimation has used file count as an indicator of project scale and structural complexity \cite{qi2017software}. Further, size of a project has been shown to have direct associations with the number of potential vulnerabilities \cite{gkortzis2021software}.

\noindent\underline{\textit{Cyclomatic Complexity and Maintainability Index of Project:}} 
We use Radon~\cite{radon}, a popular code analysis tool~\cite{liawatimena2018django, hassan2024evaluating}, to compute each project's average maintainability index and cyclomatic complexity. Cyclomatic complexity is computed as the average of complexity scores across all functions in the codebase. Higher values indicate more intricate control logic throughout the project. We examine whether project complexity is associated with vulnerability introduction.

Maintainability index is derived from several factors, including cyclomatic complexity, lines of code, and Halstead volume, and is designed to estimate how easy it is to maintain the codebase. A higher maintainability index suggests code that is easier to understand and modify. We examine whether this metric is associated with vulnerability introduction.
 
We measure the code, issue, and project-level metrics on the entire SWE-bench dataset. For Llama-generated vulnerabilities, we report the metrics calculated on the data associated with all 21,294 issues in the dataset. For GPT-5.4 nano and agent-generated vulnerabilities, these metrics are calculated only on the corresponding data associated with the \textit{test} set of 2,294 issues.
\section{Results}
\label{results}

\subsection{RQ1: Security of patches generated by Developers vs. Standalone LLMs}
\label{rq1-results}

\textbf{Vulnerability Prevalence:} Table~\ref{tab:vuln_stats} presents the number of files with new vulnerabilities introduced by \textit{Developers} and \textit{standalone LLM} patches. The raw counts vary significantly across tools and code sources. For the full SWE-bench dataset (train+test), the tools detected 1440 new vulnerability findings in Llama 3.3 patches. Of these findings, 185 (12.8\%) passed the majority-vote filter, spanning 181 distinct SWE-bench issues. Upon manual inspection, 135/185 candidates remained, spanning 131 distinct issues. In contrast, developer patches for the full dataset triggered 1398 alerts, but only 17 candidates (1\%) were retained after majority vote filtering. We manually verified 12/17 of these candidates to be valid security concerns. On the SWE-bench test set, GPT-5.4 nano produced 6 majority-vote candidates, of which 3 were retained after manual inspection. On the same test split, Llama 3.3 produced 14 majority-vote candidates and 9 retained vulnerabilities, while developer patches produced 3 majority-vote candidates and 2 retained vulnerabilities.

These retained counts should be read as manually confirmed vulnerabilities after conservative filtering, rather than as a complete measure of all vulnerabilities in APR patches. Thus, the resulting vulnerability types reflect recurring patterns in the retained patches.

\begin{table}[h]
\centering
\footnotesize
\resizebox{\linewidth}{!}{%
\begin{tabular}{p{0.22\linewidth} p{0.18\linewidth} p{0.18\linewidth} p{0.20\linewidth} p{0.20\linewidth}}
\toprule
\textbf{Patch Source} &
\textbf{Split} &
\textbf{Majority Vote} &
\textbf{Retained Vulnerabilities} &
\textbf{Affected Issue Instances} \\
\midrule
Llama 3.3 & Train + Test & 185 & 135 & 131 \\
Developers & Train + Test & 17 & 12 & 12 \\
\midrule
Llama 3.3 & Test & 14 & 9 & 8 \\
GPT-5.4 nano & Test & 6 & 3 & 3 \\
Developers & Test & 3 & 2 & 2 \\
OpenHands & Test & 195 & 44 & 11 \\
AutoCodeRover & Test & 3 & 3 & 3 \\
HoneyComb & Test & 2 & 2 & 2 \\
SWE-agent & Test & 54 & 4 & 4 \\
\bottomrule
\end{tabular}
}
\caption{Number of retained vulnerabilities and affected issue instances across evaluation splits.}
\label{tab:vuln_stats}
\end{table}

When comparing individual tools, Bandit reported the highest number of vulnerabilities overall: 1383 in Llama 3.3-generated code and 1055 in developer code. Semgrep detected 323 vulnerabilities in Llama 3.3 outputs but just 9 in developer code. CodeQL, the most conservative tool, identified 62 vulnerabilities in Llama 3.3-generated code and 41 in developer code. For GPT-5.4 nano, Bandit reported 59 new findings, Semgrep reported 12, and CodeQL reported 0, for 71 new static-analysis findings in total. Overall, our results suggest that, consistent with prior work~\cite{zhou2024comparison}, Bandit produces more false positives which had to be removed in the next two steps.

It is worth noting that all reported numbers are the new vulnerabilities that were injected only after the LLM/Developers modified the files or added the code to the repository. These vulnerabilities did not exist in the code before the patches were applied. For instance, in an issue from the \textit{Qiskit} project, Llama introduced a CWE-95 vulnerability by using \texttt{eval()} to interpret user-provided input strings as Python code. These strings are strictly sanitized in developers' code. The LLM-generated code, however, converted the string into a Python lambda function without proper sanitization or sandboxing. This enabled the possibility of a code injection attack, since malicious input could be executed directly.

\textbf{Vulnerability Types and Overlaps:}
Table~\ref{tab:llm_dev_top_cwes} contains the most common vulnerabilities in standalone LLM-generated and developer-written code \footnote{Each tool can assign multiple CWEs to the same finding and each vulnerability can be assigned different CWEs by different tools. Therefore the total number of CWE types is higher than the total number of vulnerabilities.}. The most common vulnerabilities in Llama 3.3-generated patches include command injection (CWE-78, 79 instances), eval injection (CWE-95, 39), insecure deserialization (CWE-502, 17), and path traversal (CWE-22, 14). While less frequent, other vulnerability types like incorrect permission assignment or improper authorization also appeared in the Llama 3.3 patches. For GPT-5.4 nano, the retained test-set vulnerabilities included insecure deserialization (CWE-502), improper exception handling and incorrectly resolved names (CWE-703 and CWE-706), and command or eval injection (CWE-78 and CWE-95). The full set of vulnerabilities are included in our replication package.

\begin{table}[h]
\footnotesize
\centering
\setlength{\tabcolsep}{4pt}
\begin{tabular}{p{0.75\linewidth} p{0.15\linewidth}}
\toprule
\textbf{Llama 3.3-Generated Code} & \\
\midrule
CWE-78 (OS Command Injection) & 79 \\
CWE-95 (Eval Injection) & 39 \\ 
CWE-502 (Insecure Deserialization) & 17 \\
CWE-22 (Path Traversal) & 14 \\
\midrule
\textbf{Developer-Written Code} & \\
\midrule
CWE-78 (OS Command Injection) & 4 \\ 
CWE-377 (Insecure Temporary File) & 3 \\
CWE-732 (Permission Misconfiguration) & 3 \\
CWE-22 (Path Traversal) & 3 \\
\midrule
\textbf{GPT-5.4 nano-Generated Code, Test} & \\
\midrule
CWE-78 (OS Command Injection) & 1 \\
CWE-95 (Eval Injection) & 1 \\
CWE-502 (Insecure Deserialization) & 1 \\
CWE-703 (Improper Exception Handling) & 1 \\
CWE-706 (Incorrectly Resolved Name or Reference) & 1 \\
\midrule
\textbf{Developer-Written Code, Test} & \\
\midrule
CWE-78 (OS Command Injection) & 1 \\
CWE-95 (Eval Injection) & 1 \\
CWE-502 (Insecure Deserialization) & 1 \\
\bottomrule
\end{tabular}
\caption{Most recurring CWE types found in standalone LLM-generated and developer-written patches. The full CWE distribution for the retained RQ1 vulnerabilities is provided in Appendix~\ref{app:rq1-full-cwe}.}
\label{tab:llm_dev_top_cwes}
\end{table}

Across the SWE-bench train+test set, developer-written patches showed far fewer retained vulnerabilities than Llama 3.3 patches, but some of the top categories i.e., command injection (CWE-78, 4) and path traversal (CWE-22, 3), are similar to the Llama 3.3 results. These two common CWEs (mostly related to improper input handling and insufficient sanitization when interacting with system commands or file paths) and their presence in both sets may stem from the fact that LLMs are trained on human code, inheriting common patterns, including insecure ones. On the SWE-bench test set, GPT-5.4 nano and developer patches also shared CWE categories, including CWE-78, CWE-95, and CWE-502. However, this overlap is based on CWE type rather than identical vulnerable code locations. GPT-5.4 nano and developer patches overlapped on one issue, \textit{django\_\_django-15139}, but the retained vulnerable files differed.

Despite the similarities in vulnerability patterns, certain categories were far more prominent in the Llama 3.3-generated code and extremely rare within developer-written patches. Notably, eval injection (CWE-95) was identified in 39 Llama 3.3 patches and only one developer patch. Similarly insecure deserialization (CWE-502) was repeated in 17 Llama 3.3 patches while being detected in only two developer patches. The manually inspected Llama 3.3 patches included uses of \texttt{eval()} to dynamically execute strings and insecure algorithms like MD5. In some inspected cases, Llama 3.3 replaced existing project abstractions with simpler but less secure implementations. For instance, in an issue requesting improved password hashing, Llama 3.3 generated a new custom hasher that used SHA-256 instead of upgrading the existing Argon2-based implementation. While SHA-256 is straightforward to apply, it lacks protective features against brute-force attacks (CWE-327). Additionally, Llama 3.3 produced the following implementation of the \texttt{check\_password()} function, which includes a vulnerable use of \texttt{hashlib.md5()}:

\begin{lstlisting}[language=Python]
def check_password(raw_password, encoded_password, setter=None):
    ...
    data = raw_password.encode('utf-8')
    return hashlib.md5(data).hexdigest() == encoded_2['hash']
\end{lstlisting}

If such patches were merged without security checks e.g., CI-integrated static analysis runs or manual reviews, the consequences for a real system could be severe. In the password-hashing example, the model introduces a secondary MD5-based check that directly compares attacker-controllable input against stored hashes. In a production environment, this would expose users to significantly higher risk of credential compromise; brute-force and dictionary attacks become dramatically more feasible, and any leakage of the hash database (through a separate vulnerability) would have far more damaging consequences than under the original design. More broadly, many of the CWE categories we observe in LLM patches e.g., eval injection, insecure deserialization, and command injection, are associated with high-impact outcomes such as arbitrary code execution, privilege escalation, or unauthorized access to sensitive data. Because these vulnerabilities are introduced in the course of fixing issues, they are unlikely to be caught by functional tests alone, and can remain latent until exploited. As a result, if LLM-generated fixes are integrated without additional security checks, they may fix the immediate bug while degrading the system’s security.

\textbf{RQ1 Summary:} Across the SWE-bench train+test set, Llama 3.3 introduced 135 manually confirmed vulnerabilities, compared to 12 in developer patches (Table~\ref{tab:vuln_stats}). On the SWE-bench test set, GPT-5.4 nano introduced 3 manually confirmed vulnerabilities, compared to 2 in developer patches. The most frequent Llama vulnerabilities include command injection (CWE-78), eval injection (CWE-95), insecure deserialization (CWE-502), and path traversal (CWE-22). GPT-5.4 nano also introduced vulnerabilities in overlapping categories, including CWE-78, CWE-95, and CWE-502. These results show that standalone LLM-generated fixes can introduce security weaknesses that differ from developer-written patches.

\subsection{RQ2: Security of patches generated by the agentic APR workflows}

\textbf{Vulnerability Prevalence:} Table \ref{tab:vuln_stats} also reports the number of vulnerabilities for agentic frameworks i.e., OpenHands, AutoCodeRover, HoneyComb, and SWE-agent when evaluated on the SWE-bench test set. Similar to RQ1, we manually inspect all instances to validate our data. 

The static analyses and the majority voting process resulted in 195, 3, 2, and 54 instances marked as potential vulnerabilities by OpenHands, AutoCodeRover, HoneyComb, and SWE-agent, respectively. As with Llama and developer patches, Bandit consistently reported the highest number of vulnerabilities for OpenHands, AutoCodeRover, and HoneyComb, followed by Semgrep and then CodeQL. For SWE-agent, Semgrep reported the highest number of raw findings, with 665 findings, followed by Bandit with 375 and CodeQL with 17. However, despite the high number of raw detections, the majority vote strategy proved highly selective: for example, OpenHands generated over 1,383 alerts across all tools cumulatively, yet only 195 vulnerabilities (approximately 15\%) were retained in the final filtered set. Similarly, SWE-agent generated 1,057 raw findings across tools, but only 54 passed majority voting. This reinforces the importance of using a stricter intersection criterion when analyzing code security. Next, led by the manual inspection process, 44 out of 195 OpenHands vulnerabilities, all 5 vulnerabilities from AutoCodeRover and HoneyComb, and 4 out of 54 SWE-agent vulnerabilities were confirmed as valid security threats.

While all four frameworks introduced new vulnerabilities, OpenHands produced a disproportionately larger number. To better understand this disparity, we closely examined the results from OpenHands. Among other factors, OpenHands is given full autonomy, not only to inspect and modify project files, but also to execute commands, run tests, and generate new scripts as part of its process. In the inspected OpenHands cases, patches sometimes included changes beyond those required by the developer-written repair. Our analysis indicates that out of the 195 vulnerabilities initially attributed to OpenHands, 12 were located in test files and 35 were found in files intended to reproduce the issue (e.g., \texttt{reproduce.py}, \texttt{reproduce\_error.py}). Additionally, 75 marked instances came from auxiliary package or tool artifacts that OpenHands placed in the repository during its execution. These files were not part of the intended project-level repair, even though they appeared in the patched repository state. We therefore exclude them from the remaining analysis to avoid attributing vulnerabilities in external or temporary artifacts to the repaired project code. Furthermore, the 44 retained OpenHands vulnerabilities span 11 affected issue instances, as reported in Table~\ref{tab:vuln_stats}. In fact, three issues alone, \textit{django ticket \#14056}, \textit{pytest issue \#5787}, and \textit{django ticket \#14681}, accounted for 35 of the 44 final vulnerabilities. In the most extreme case, \textit{django ticket \#14056}, OpenHands generated a diff file that exceeded 500,000 lines in length. In these inspected OpenHands cases, we observed extensive file rewrites or reimplementation of existing core software components. The concentration of retained findings in a small number of issues limits generalizability, and our analysis does not isolate the framework's autonomy or control mechanisms as the cause of these outcomes.

In one instance, OpenHands attempted to resolve an issue in the \textit{django} project, related to invalid SQL generation from constraint expressions involving ordering (e.g., \texttt{Lower("name").desc()}). The original bug caused form validation to fail with SQL syntax errors because \texttt{DESC} was incorrectly placed inside a \texttt{WHERE} clause. The developer patch resolved this cleanly by removing any ordering operations (like \texttt{desc()}) before constructing validation expressions:

\begin{lstlisting}[language=Python]
if isinstance(expr, OrderBy):
    expr = expr.expression
expressions.append(Exact(expr, expr.replace_expressions(...)))
\end{lstlisting}

In contrast, the OpenHands' patch bypassed Django's ORM entirely and rewrote the validation logic using raw SQL:

\begin{lstlisting}[language=Python]
where_clause = " AND ".join(conditions)
sql = f"SELECT 1 FROM {table_name} WHERE {where_clause} LIMIT 1"
cursor.execute(sql, params)
\end{lstlisting}

While this approach can address the original issue, it introduced a new vulnerability,  CWE-89 (SQL injection), flagged by both Bandit and Semgrep. Specifically, although query parameters were bound securely, the \texttt{field\_name} values, used to construct column identifiers, were directly inserted into the query string:
\begin{lstlisting}[language=Python]
conditions.append(f"{qn(field_name)} = %s")
\end{lstlisting}

If these expressions contain malicious input (e.g., \texttt{F("name); DROP TABLE users; --")}), they could directly manipulate query. Django’s ORM normally safeguards against such risks, but manual SQL construction ignores these protections. This example illustrates how a model, may introduce security risks by deviating from design principles.

\textbf{Vulnerability Types and Overlaps:} Among the agentic frameworks, OpenHands, responsible for 44 vulnerabilities, exhibited a similar vulnerability pattern to that of Llama 3.3: command injection (CWE-78, n=10), eval injection (CWE-95, n=8), and insecure deserialization (CWE-502, n=6) were all among the common CWEs in OpenHands patches. This alignment suggests that even when LLMs are embedded within frameworks, some core security risks and patterns persist.

Interestingly, even the relatively smaller number of vulnerabilities generated by AutoCodeRover and HoneyComb included categories like command injection and eval injection. SWE-agent showed a similar pattern; its retained vulnerabilities included command injection (CWE-78, n=2), eval injection (CWE-95, n=2), insecure deserialization (CWE-502, n=2), and improper exception handling (CWE-703, n=1). Given that the dataset used in RQ2 is smaller than in RQ1, the presence of these CWE types across all four agentic frameworks highlights the persistent security risks posed by autonomous repair agents.

Moreover, we find that vulnerabilities introduced by the agentic frameworks overlap with developer-written patches at the CWE level, but not at the exact code-location level. Developer patches on the SWE-bench test set included CWE-78, CWE-95, and CWE-502. OpenHands and SWE-agent shared all three of these CWE categories with developer patches, while AutoCodeRover and HoneyComb shared CWE-78 and CWE-95. However, none of the agentic frameworks overlapped with developers on the same issue, same file, and same CWE. These results echo the same observations in Section \ref{rq1-results}. While developers, Llama, and agentic frameworks all can add vulnerabilities to the codebase, the vulnerabilities produced by the LLMs or agentic frameworks are not simply repeating the developers' mistakes.

\textbf{RQ2 Summary:} While the agentic frameworks were evaluated on a smaller subset of data (test-set), \textbf{all four still introduced manually confirmed vulnerabilities}. OpenHands produced 44 retained vulnerabilities across 11 affected issue instances, compared to 3 vulnerabilities across 3 issue instances for AutoCodeRover, 2 across 2 for HoneyComb, and 4 across 4 for SWE-agent. The retained counts differ across the evaluated frameworks, with OpenHands findings concentrated in a small number of issues. These observations do not establish that greater autonomy caused the differences.

\subsection{RQ3: What code, issue, or project characteristics are associated with the generation of vulnerable code by LLMs and agentic APR frameworks when resolving issues?}
\label{rq23-results}

\subsubsection{Code-Level Factors}
We analyze the code-level characteristics of the files associated with vulnerability-inducing LLM and agentic APR patches and compare them to those from all modified or generated files. Table~\ref{tab:code-metrics-summary} summarizes these comparisons for standalone LLM-generated patches. We note that the dataset for Llama 3.3 is significantly larger than the test-set analyses. As a result, we treat the GPT-5.4 nano, AutoCodeRover, HoneyComb, and SWE-agent comparisons as descriptive because their retained vulnerability counts are small. For agentic frameworks, we include the raw values for all four frameworks in Table~\ref{tab:agentic-code-metrics-comparison}.

\begin{table*}[t]
\centering
\footnotesize
\setlength{\tabcolsep}{4pt}
\renewcommand{\arraystretch}{0.92}
\begin{tabular*}{\textwidth}{@{\extracolsep{\fill}}lrrrr@{}}
\toprule
\textbf{Source / Metric} & \textbf{Vuln Mean} & \textbf{All Mean} & \textbf{p-value} & \textbf{Cliff's $\delta$} \\
\midrule
\multicolumn{5}{@{}l}{\textbf{Llama 3.3}} \\
Files Added or Modified (All) & 1.72 & 1.25 & $<$0.001 & 0.200 \\
Files Added or Modified (Python) & 1.60 & 1.16 & $<$0.001 & 0.199 \\
Unique File Types Modified & 1.10 & 0.99 & 0.007 & 0.081 \\
Sensitive Files Modified & 0.00 & 0.007 & 0.369 & -0.006 \\
LOC Generated by LLM (Python) & 1740.50 & 2218.12 & $<$0.001 & -0.197 \\
Avg. Cyclomatic Complexity & 2.88 & 3.02 & 0.8796 & -0.009 \\
\midrule
\multicolumn{5}{@{}l}{\textbf{GPT-5.4 nano}} \\
Files Added or Modified (All) & 2.00 & 0.98 & 0.104 & 0.363 \\
Files Added or Modified (Python) & 2.00 & 0.97 & 0.104 & 0.368 \\
Unique File Types Modified & 1.00 & 0.89 & 0.537 & 0.114 \\
Sensitive Files Modified & 0.00 & 0.00 & 1.000 & 0.000 \\
LOC Generated by LLM (Python) & 320.00 & 523.71 & 0.400 & -0.281 \\
Avg. Cyclomatic Complexity & 4.21 & 3.65 & 0.603 & 0.173 \\
\bottomrule
\end{tabular*}
\caption{Summary statistics comparing vulnerable vs. all standalone LLM-generated code for code-level metrics.}
\label{tab:code-metrics-summary}
\end{table*}

\begin{table}[h]
\centering
\footnotesize
\resizebox{\linewidth}{!}{%
\begin{tabular}{lrrrrrrrr}
\toprule
\textbf{Metric} &
\multicolumn{2}{c}{\textbf{OpenHands}} &
\multicolumn{2}{c}{\textbf{AutoCodeRover}} &
\multicolumn{2}{c}{\textbf{HoneyComb}} &
\multicolumn{2}{c}{\textbf{SWE-agent}} \\
 & \textbf{V} & \textbf{A} & \textbf{V} & \textbf{A} & \textbf{V} & \textbf{A} & \textbf{V} & \textbf{A} \\
\midrule
Files Added or Modified (All) & 344.27\textsuperscript{*} & 11.38 & 1.33 & 1.07 & 1.50 & 1.46 & 5.75 & 3.02 \\
Files Added or Modified (Python) & 68.83\textsuperscript{*} & 4.46 & 1.33 & 1.07 & 1.50 & 1.40 & 5.25 & 1.99 \\
Unique File Types Modified & 3.73\textsuperscript{*} & 1.41 & 1.00 & 0.98 & 1.00 & 0.90 & 1.50 & 1.08 \\
Sensitive Files Modified & 0.0909\textsuperscript{*} & 0.0136 & 0.00 & 0.00 & 0.00 & 0.004 & 0.00 & 0.004 \\
LOC Generated by LLM (Python) & 1218.6 & 1591.0 & 387.33 & 1294.6 & 1591 & 1764 & 2104.0 & 1615.98 \\
\makecell[l]{Avg. Cyclomatic Complexity\\of Generated Code} & 4.015 & 4.513 & 7.352 & 4.303 & 6.470 & 4.363 & 3.80 & 4.41 \\
\bottomrule
\end{tabular}
}
\caption{Mean values for vulnerable (V) and all (A) agentic-generated patches across code-level metrics. \textsuperscript{*} marks OpenHands metrics with statistically significant vulnerable versus all differences. AutoCodeRover, HoneyComb, and SWE-agent are interpreted descriptively because they have 3, 2, and 4 retained vulnerable instances, respectively.}
\label{tab:agentic-code-metrics-comparison}
\end{table}

\noindent\underline{ \textit{\# Files Added or Modified:}} As shown in Table~\ref{tab:code-metrics-summary}, Llama 3.3-generated vulnerable patches (\textit{Vuln Mean}) involve modifications to a larger number of files compared to all Llama 3.3-generated patches \textit{(All Mean)}, both overall (1.72 vs. 1.25) and within Python files (1.60 vs. 1.16), with statistically significant differences (p $<$ 0.001, Cliff's $\delta$ = 0.2 \& 0.199). GPT-5.4 nano shows the same descriptive direction on the test set: vulnerable patches modified more files overall (2.00 vs. 0.98) and more Python files (2.00 vs. 0.97), although the small number of retained vulnerabilities makes this comparison descriptive rather than conclusive.

Among the agentic frameworks, OpenHands showed an even more dramatic pattern: vulnerable instances modified 344.27 files on average (vs. 11.38), including Python files (68.83 vs. 4.46), with large effect size of $\delta$=0.77 and a p-value$<$0.001. This reflects broad edits among the 11 affected OpenHands issue instances and should not be generalized to all autonomous APR systems. Additionally, the differences in AutoCodeRover (1.33 vs. 1.07) and HoneyComb (1.50 vs. 1.46) showed higher means for vulnerable instances. SWE-agent follows the same direction, with vulnerable patches modifying more files overall (5.75 vs. 3.02) and more Python files (5.25 vs. 1.99).

Figure~\ref{fig:files-changed-boxplot} compares the distribution of changed/added Python files per instance between vulnerable and all patches, across sources. We observe that all the tools often show a general increase in the number of modified files in vulnerable instances relative to the rest. OpenHands stands out with a substantial shift, indicating significantly broader edits. Llama-generated patches, similar to AutoCodeRover and HoneyComb, also display strong differences with lower overall file counts. These trends highlight that vulnerability-prone patches tend to involve more scattered edits across multiple files.

\noindent\underline{\textit{\# Unique File Types Modified:}} Vulnerable code modifications by Llama affected a slightly broader range of file types (mean = 1.10 vs. 0.99). This difference is statistically significant (p = 0.007), although the effect size is negligible (Cliff's $\delta$ = 0.081). This suggests that file type diversity alone is a weak indicator of vulnerability risk but may still inform broader risk assessments when combined with other features. GPT-5.4 nano shows a similar descriptive pattern, with vulnerable patches touching slightly more unique file types than all generated patches (1.00 vs. 0.89).

Across the frameworks, we observe similar patterns: AutoCodeRover (1.00 vs. 0.98) and HoneyComb (1.00 vs. 0.90) both show higher mean values for the vulnerable instances. Further, OpenHands exhibited a more notable increase (3.73 vs. 1.41), with statistical significance and meaningful effect size ($\delta$ = 0.345, p = 0.0011). SWE-agent also shows a higher mean for vulnerable patches (1.50 vs. 1.08). This reinforces earlier findings that broader file editing behavior is associated with vulnerable generated patches.

\begin{figure*}[t]
    \centering
    \includegraphics[width=\textwidth]{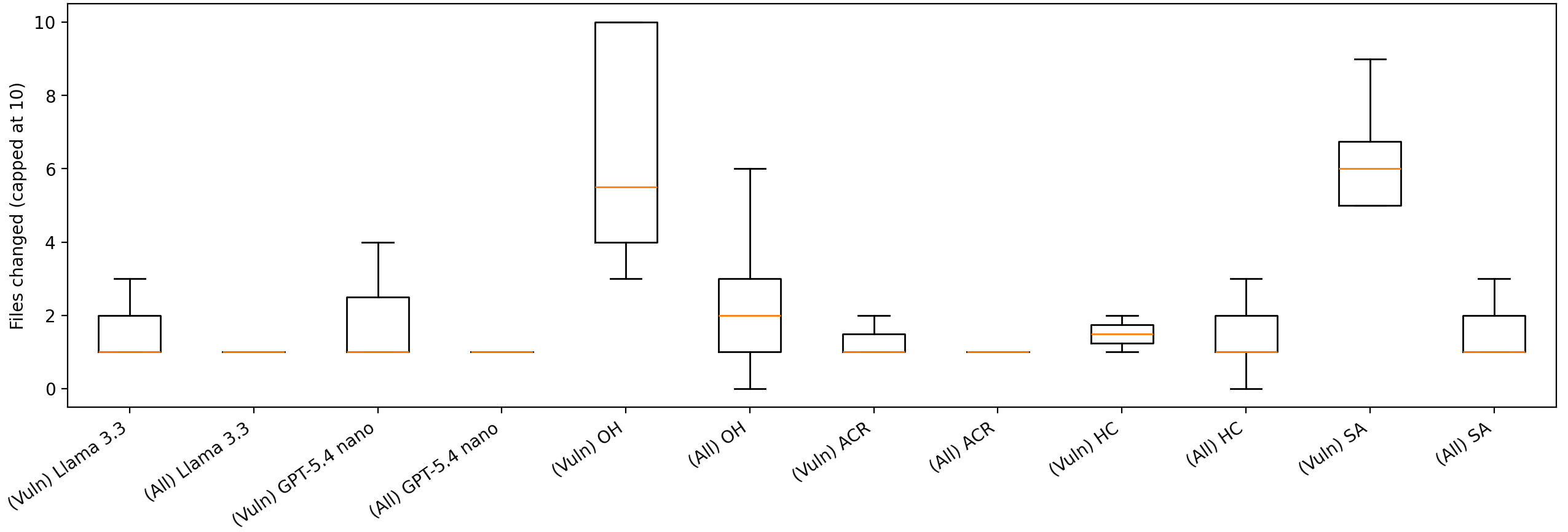}
    \vspace{-1em}
    \caption{Number of changed Python files per instance across standalone LLMs and agentic APR frameworks, capped at 10.}
    \label{fig:files-changed-boxplot}
\end{figure*}

\noindent\underline{\textit{\# Sensitive file modifications:}} Using our predefined set of sensitive file types (e.g., \texttt{.conf}, \texttt{.yml}, \texttt{.json}), we compare their modification rates in vulnerable and all instances. In Llama-generated code, vulnerable instances did not show significant changes in modifying such files (p = 0.369), indicating that Llama 3.3 does not disproportionately touch sensitive configuration files when generating insecure code. GPT-5.4 nano also did not show sensitive file modifications in the retained vulnerable cases (0.00 vs. 0.00).

This pattern holds for AutoCodeRover (0.0 vs. 0.0) and HoneyComb (0.0 vs. 0.004). SWE-agent also follows this pattern, with no sensitive file modifications in vulnerable patches (0.00 vs. 0.004). OpenHands, on the other hand, exhibited a statistically significant increase in sensitive file edits (0.09 vs. 0.01, p $<$ 0.008), with sensitive file modifications occurring more frequently in its vulnerable patches. However, the effect size remained negligible ($\delta$ = 0.081), and results across frameworks were mixed. Sensitive file edits were associated with vulnerable patches in OpenHands, but this association was not consistent across frameworks.

\noindent\underline{\textit{LOC Generated by LLM:}} This factor represents the scale of code generation by the LLM. In the Llama-generated patches, vulnerable instances involve less extensive edits (1740 vs. 2218 lines on average), a statistically significant difference (p $<$ 0.001) with a negative effect size ($\delta$ = - 0.197). GPT-5.4 nano shows the same descriptive direction on the test set, with vulnerable patches generating fewer Python LOC on average (320.00 vs. 523.71). This suggests that larger code generations are not necessarily more prone to vulnerabilities. As demonstrated in prior work, even highly concise edits can introduce substantial security risks. For instance, severe vulnerabilities such as Heartbleed and ShellShock stemmed from only a handful of modified lines, showing how small changes to core logic can have extreme consequences~\cite{lovgren2023can}. This finding also aligns with empirical studies of developer activity, revealing that vulnerability-inducing commits are shrinking in size over time, with smaller commits associated with more severe vulnerabilities~\cite{andrade2021investigating, jiang2024understanding}. These LOC differences describe the extent of generated changes, rather than establishing how the models reasoned about program context.

Among the agentic frameworks, AutoCodeRover shows the strongest contrast, with vulnerable outputs averaging fewer lines (387 vs. 1294). OpenHands also shows a reduction in generated LOC for vulnerable instances (1218 vs. 1591), with a small effect size ($\delta$ = -0.10, p = 0.51). HoneyComb also shows lower mean Python file LOC for vulnerable cases (1591 vs. 1764). In contrast, SWE-agent shows higher mean Python file LOC for vulnerable cases than across all SWE-agent patches (2104.0 vs. 1615.98). These contrasting patterns suggest that while the scale of generation can be associated with vulnerability likelihood, the relationship is highly dependent on the specific framework's behavior and other contextual factors.

\noindent\underline{\textit{Cyclomatic Complexity of Generated Code:}} We assess whether the vulnerable outputs contain more complex control flows by measuring the average cyclomatic complexity of functions in files generated or modified by the LLM or agentic APR framework. In the Llama patches, vulnerable instances show slightly lower complexity (2.88 vs. 3.02), but this difference is not statistically significant (p = 0.8796, $\delta$ = -0.009), suggesting that structurally simple code may still contain security flaws. GPT-5.4 nano shows a descriptive increase in average cyclomatic complexity for vulnerable patches (4.21 vs. 3.65), but the comparison is not statistically significant.

Among agentic frameworks, OpenHands shows a similar trend, with vulnerable files having lower complexity (4.015 vs. 4.513), though the difference is not statistically significant (p = 0.688, $\delta$ = 0.035). Further, HoneyComb shows a slight increase in the complexity of vulnerable files (6.47 vs. 4.363). AutoCodeRover exhibits a notable increase in complexity of vulnerable files (7.352 vs. 4.303). SWE-agent follows the lower-complexity direction, with vulnerable patches averaging 3.80 compared to 4.41 across all SWE-agent patches. Our observations highlight that vulnerability is not limited to structurally complex code, as simple logic can still be insecure. Yet, abnormally high complexity, as with the patches of AutoCodeRover, can serve as an indicator for security risk, offering a practical signal for prioritizing code review.

\vspace{-.5em}
\begin{tcolorbox}[colback=gray!5, colframe=gray!80, title=\textbf{Code-Level Factors}, boxsep=2pt, left=4pt, right=4pt, top=2pt, bottom=2pt] 
Across systems, vulnerable patches tend to modify more files. File-type diversity, generated LOC, and cyclomatic complexity show less consistent patterns across systems, suggesting that these factors are useful context but not stable predictors on their own.
\end{tcolorbox}
\vspace{-.5em}

\subsubsection{Issue-Level Factors}
We analyze the characteristics of GitHub issues to determine whether specific issue features correlate with the introduction of vulnerabilities. For GPT-5.4 nano and SWE-agent, we report issue-level characteristics descriptively because they have only 3 and 4 retained vulnerable instances, respectively.

We count each distinct SWE-bench issue associated with a retained vulnerable patch once. For Llama 3.3, 101 of the 131 affected issues (77.1\%) were classified as \textit{Bug}, 29 (22.1\%) as \textit{Feature}, and 1 (0.8\%) as \textit{Question}. OpenHands had 6 bug issues and 5 feature issues; AutoCodeRover had 2 bug issues and 1 feature issue; and HoneyComb had 2 bug issues. For GPT-5.4 nano, 2 of the 3 retained vulnerable instances were labeled as \textit{Bug}, and 1 was labeled as \textit{Feature}. For SWE-agent, 3 of the 4 retained vulnerable instances were labeled as \textit{Bug}, and 1 was labeled as \textit{Feature}. Here, OpenHands stands out as it often generated vulnerable code when attempting to resolve issues that seek to implement new features within projects. The feature-labeled GPT-5.4 nano and SWE-agent instance is \textit{django\_\_django-15139}, which we classify as a security-related feature or deprecation request rather than a standard functional bug report.

Table \ref{tab:bug type} summarizes the distribution of bug types across all retained vulnerable instances. The most frequent categories are \textit{Program Anomaly} and \textit{Configuration}. Other bug types, such as \textit{Performance} and \textit{GUI-related}, appear less frequently. The newly annotated GPT-5.4 nano and SWE-agent instances follow a similar pattern, with most bug-labeled cases falling under \textit{Program Anomaly}. They also include security or deprecation-related cases. Overall, vulnerabilities of Llama 3.3 and the agentic frameworks mostly involve addressing core logic or environment-related issues, where deeper contextual understanding is critical.

\begin{table*}[t]
\centering
\footnotesize
\setlength{\tabcolsep}{4pt}
\renewcommand{\arraystretch}{0.95}
\begin{tabular*}{\textwidth}{@{\extracolsep{\fill}}lcccccc@{}}
\toprule
\textbf{Bug Type} & \textbf{Llama} & \textbf{OpenHands} & \textbf{AutoCodeRover} & \textbf{HoneyComb} & \textbf{GPT} & \textbf{SWE-agent} \\
\midrule
Program Anomaly Issue           & 59 & 5 & 1 & 2 & 2 & 2 \\
Configuration Issue             & 18 & 1 & -- & -- & -- & -- \\
Miscellaneous/Other Issue       & 8 & -- & -- & -- & -- & -- \\
Performance Issue               & 5 & -- & -- & -- & -- & -- \\
GUI-Related Issue               & 3 & -- & -- & -- & -- & -- \\
Test Code-Related Issue         & 3 & -- & -- & -- & -- & -- \\
Network Issue                   & 3 & -- & -- & -- & -- & -- \\
Permission/Deprecation Issue    & 2 & -- & -- & -- & -- & 1 \\
Security Issue                  & -- & -- & 1 & -- & 1 & 1 \\
\bottomrule
\end{tabular*}
\caption{Bug type distribution across vulnerable instances.}
\label{tab:bug type}
\end{table*}

\noindent\underline{\textit{Information Completeness:}} Among the 101 distinct bug-related issues associated with retained vulnerable Llama 3.3 patches, 97 (96.0\%) included \textit{observed behavior}, 66 (65.3\%) included \textit{steps to reproduce}, and 40 (39.6\%) provided \textit{expected behavior}. These rates indicate that while problems are usually described, some important information, i.e., what the code is supposed to do and how to reproduce the issue, is often missing. Missing expected behavior and reproduction steps were recurring contextual characteristics among these vulnerable instances.

We examined 10 distinct bug-related issues across OpenHands, AutoCodeRover, and HoneyComb, comprising 6, 2, and 2 issues, respectively. All 6 OpenHands issues included observed behavior, expected behavior, and reproduction steps. Both AutoCodeRover issues included observed behavior and reproduction steps, while 1 included expected behavior. Both HoneyComb issues included all three types of information. For GPT-5.4 nano, all 3 retained vulnerable instances included observed behavior, 2 included expected behavior, and 2 included steps to reproduce. For SWE-agent, all 4 retained vulnerable instances included observed and expected behavior, while 2 included steps to reproduce. While sample sizes are limited, we observe that framework-generated vulnerabilities can occur even when issue descriptions are reasonably specified.

\noindent\underline{\textit{\# Words:}} We observed minor differences in the length of issue descriptions and associated comments in vulnerable vs. all Llama-generated instances. On average, problem descriptions contained 111.71 words for all issues and 105.78 words for vulnerable instances. Neither the word count of problem descriptions alone (p = 0.5346, Cliff's $\delta$ = -0.031) nor the combined word count including comments (p = 0.4726, $\delta$ = 0.036) reached statistical significance. These results show no statistically significant association between issue-content length and vulnerability occurrence in the Llama 3.3 comparison. Other aspects of issue information, such as clarity or specificity, warrant further investigation.

Across the frameworks, we observed mixed patterns. OpenHands and AutoCodeRover had higher average word counts when combining issue and comment text (547.16 and 435.67 words, respectively), while HoneyComb's was much lower (96.0). GPT-5.4 nano retained vulnerable instances had 117.33 issue-body words on average and 126.00 words when comments were included. SWE-agent showed a much higher average, with 371.00 issue-body words and 1696.50 words when comments were included, largely due to one highly discussed Astropy issue. These values describe framework-specific differences in the issue information associated with the retained vulnerable patches.

\noindent\underline{\textit{\# Comments Prior to First PR:}} Llama 3.3-generated vulnerable instances had slightly more comments on average (4.1) compared to all issues (3.6), but the difference was not significant (p = 0.68, $\delta$ = 0.025), indicating that comment count does not clearly distinguish vulnerable from non-vulnerable Llama 3.3-generated patches.

Across OpenHands, AutoCodeRover, and HoneyComb, none of the vulnerable issues contained comments prior to the first pull request. GPT-5.4 nano had 0.33 comments before the first PR on average, with only one retained vulnerable instance having a prior comment. SWE-agent had 2.75 comments before the first PR on average, again driven by one Astropy issue with 11 prior comments. This suggests that prior discussion is not consistently associated with the vulnerabilities observed within the frameworks.

\noindent\underline{\textit{Presence of Code Snippets:}}  
We observe that only 47.4\% of issues associated with retained vulnerable Llama 3.3 outputs include code snippets in the issue body, compared to 58.6\% of all issues. This difference is statistically significant ($\chi^2$ = 6.44, p = 0.011), suggesting that the presence or absence of code snippets is not evenly distributed between the two groups. The chi-squared test evaluates whether there is a meaningful association between two categorical variables, in this case, code snippet presence and vulnerability occurrence.
To assess the strength of this association, we calculate the phi coefficient ($\phi$ = 0.017), a standardized effect size for 2$\times$2 contingency tables. While statistically significant, the effect size is negligible, indicating a weak association. In this evaluated setting, vulnerable Llama 3.3 patches were more often associated with issue descriptions lacking code snippets, although the association was weak. For GPT-5.4 nano, 2 of the 3 retained vulnerable instances had code snippets in the issue body, and none had code snippets in the comments. For SWE-agent, 2 of the 4 retained vulnerable instances had code snippets in the issue body, and 1 had code snippets in the comments. Absent code snippets were associated with vulnerable Llama 3.3 patches in the evaluated setting, while vulnerabilities also occurred in the small GPT-5.4 nano and SWE-agent samples with code snippets.

\vspace{-.5em}
\begin{tcolorbox}[colback=gray!5, colframe=gray!80, title=\textbf{Issue-Level Factors}, boxsep=2pt, left=4pt, right=4pt, top=2pt, bottom=2pt] 
Insecure patches are often associated with bug reports and core logic issues, but the newly annotated GPT-5.4 nano and SWE-agent cases also include security or deprecation-related feature requests. Missing expected behavior, missing reproduction steps, and absent code snippets were observed among vulnerable instances, although vulnerabilities also occurred when issue reports were relatively complete.
\end{tcolorbox}
\vspace{-.5em}

\subsubsection{Project-Level Factors}
We also examined whether broader project characteristics, such as codebase size, complexity, and contributor count, are associated with the number of vulnerabilities introduced by LLM or agent-generated code. We used Spearman's rank correlation coefficient ($\rho$) to assess monotonic relationships between each project-level factor and the number of detected vulnerabilities across repositories.

\noindent\underline{\textit{\# Files in Repository:}}
We examined whether the size of the codebase, as measured by the total number of Python files in the repository, correlates with vulnerability outcomes. For Llama-generated code, the correlation was weak and not statistically significant ($\rho = 0.136$, p = 0.362). The GPT-5.4 nano test-set results also showed no statistically significant correlation ($\rho = 0.473$, p = 0.120). Similarly, we found no meaningful correlations in any of the agentic frameworks: OpenHands ($\rho = 0.168$, p = 0.602), AutoCodeRover ($\rho = 0.355$, p = 0.258), and HoneyComb ($\rho = 0.194$, p = 0.545). SWE-agent followed the same pattern ($\rho = 0.473$, p = 0.120). These findings suggest that codebase size is not a reliable indicator of LLM-generated vulnerabilities.

\noindent\underline{\textit{Maintainability and Complexity:}}
Similarly, we found no statistically significant correlation between vulnerability counts and average maintainability index or the cyclomatic complexity per function in the projects. Within the Llama 3.3-generated code, maintainability ($\rho = -0.006$, p = 0.968) and complexity ($\rho = -0.062$, p = 0.679) were not correlated with vulnerability introduction. GPT-5.4 nano showed the same pattern for maintainability ($\rho = 0.139$, p = 0.666) and complexity ($\rho = 0.084$, p = 0.796). The same holds for the frameworks. For example, in OpenHands, maintainability ($\rho = -0.021$, p = 0.888) and complexity ($\rho = 0.049$, p = 0.744) showed no statistically significant correlation with vulnerability counts. AutoCodeRover and HoneyComb showed similar patterns. SWE-agent also showed no statistically significant correlation for maintainability ($\rho = 0.108$, p = 0.739) or complexity ($\rho = -0.178$, p = 0.581). These findings show no statistically significant correlation between project complexity or maintainability and vulnerability counts in the evaluated systems.

\noindent\underline{\textit{Contributor Count:}}  
We also examined whether the number of contributors in a project correlates with the introduction of vulnerabilities. For Llama-generated code, the association was weak and statistically insignificant ($\rho = 0.184$, p = 0.217). GPT-5.4 nano also showed no statistically significant association ($\rho = 0.418$, p = 0.176). The same held true across all agentic frameworks: OpenHands ($\rho = 0.210$, p = 0.156), AutoCodeRover ($\rho = 0.158$, p = 0.288), and HoneyComb ($\rho = 0.109$, p = 0.467). SWE-agent followed the same pattern ($\rho = 0.108$, p = 0.739). These findings suggest that community size, used here as a rough proxy for project activity, showed no statistically significant association with vulnerability outcomes in the evaluated systems, although the consistently positive (albeit weak) correlations hint at a very slight directional trend.

\vspace{-.5em}
\begin{tcolorbox}[colback=gray!5, colframe=gray!80, title=\textbf{Project-Level Factors}, boxsep=2pt, left=4pt, right=4pt, top=2pt, bottom=2pt] 
Project size, complexity, and contributor count showed weak, non-significant correlations with vulnerability rates. This pattern also held for GPT-5.4 nano and SWE-agent. These factors showed no statistically significant association with vulnerability counts in the evaluated systems.
\end{tcolorbox}
\vspace{-.5em}

\noindent\textbf{RQ3 Summary:} Across all systems, vulnerable patches are most strongly associated with code-level characteristics rather than project-level properties. At the code level, insecure patches consistently modify more files. However, file-type diversity, generated LOC, and cyclomatic complexity show mixed patterns after adding GPT-5.4 nano and SWE-agent, so they are not stable predictors on their own. At the issue level, our manually annotated issue-level set shows that vulnerabilities often appear in bug reports and configuration-related issues, especially when expected behavior and reproduction steps are under-specified and when issue descriptions lack concrete code snippets. In contrast, project-level factors such as repository size, cyclomatic complexity, and contributor count show non-significant correlations with vulnerability counts.
\section{Threats to Validity}

Our security analysis uses static analysis tools, which can produce false positives and vary in coverage. To mitigate this threat, we used three tools (Semgrep, Bandit, and CodeQL), maximizing coverage, and applied a majority voting technique to reduce false positives and increase confidence in the identified vulnerabilities. We further performed a manual analysis on all majority-vote candidates to remove remaining false positives. However, our results may still miss vulnerabilities that are not detected by the static analysis tools or that are detected by only one tool and removed by majority voting. Vulnerabilities may remain undetected when tools do not support the relevant APIs or vulnerability types, or when their analysis models do not capture the relevant program behavior. Thus, our reported vulnerability counts should be interpreted as conservative estimates rather than complete measurements of all security defects.Larger or more heavily modified files may have more opportunities to receive warnings from multiple tools and satisfy file-level voting. Although manual validation removes incorrect findings within selected files but does not eliminate this selection effect, which may affect the correlations reported in RQ3.

To assess whether majority voting excludes potentially relevant security findings, we manually reviewed a sample of 60 discarded single-tool warnings from Bandit, Semgrep, and CodeQL, including 20 warnings from each tool. Using the assessment criterion described in Section \ref{methodology}, we identified six warnings associated with newly added insecure code patterns across five patches. Four warnings were associated with developer-written patches, and two were from AI-generated patches. We excluded pre-existing insecure code patterns and warnings in test or dependency artifacts, as these did not represent newly introduced insecure code. These findings confirm that majority voting can exclude relevant security findings, highlighting a tradeoff between reducing the manual validation effort and retaining potentially vulnerable code patterns. However, given the exploratory nature and limited size of our sample for manual analysis on discarded warnings, these observations should not be interpreted as estimates of overall false-negative rates.

Moreover, our evaluation focused on two standalone LLMs and four publicly available agentic APR frameworks. While there are many other models and frameworks in the space, our selection captures a relevant but not exhaustive sample of current LLM-based repair systems. Additionally, while our Llama 3.3 approach generated responses for both the train and test sets (a process that took well over a week), we acknowledge the limitation in terms of data scale and have adjusted our conclusions accordingly. GPT-5.4 nano and the agentic APR frameworks are evaluated on the SWE-bench test set only. Therefore, we avoid direct statistical comparisons between the train+test standalone LLM setting and the test-set framework setting. For instance, AutoCodeRover reports a cost of \$0.70 per instance to achieve their test set results, meaning it would cost over \$13,000 to generate patches for the full 19,000 instances of training set, which is beyond our available resources. To keep our conclusions accurate, we therefore refrain from making strong statistical claims about the agentic systems, GPT-5.4 nano, or cross-setting comparisons; report their raw counts transparently; focus on qualitative analysis of the vulnerable instances they produce; and avoid direct quantitative comparisons with the Llama 3.3 train+test results.

Further, our results reflect the combined behavior of each model, prompt, context, decoding setup, and patch-generation pipeline. While known benchmark-quality issues may influence patch generation and limit generalizability, they do not necessarily invalidate the identified security vulnerabilities in our study. We use standard SWE-bench prompts to match the benchmark setting and keep the standalone LLM evaluation consistent. However, we do not isolate the effect of prompt design from the model itself. Security-specific prompts or different generation strategies may change both the frequency and types of vulnerabilities produced. A systematic prompt ablation is outside the scope of this study, but it is an important direction for future work.

We also acknowledge that our RQ3 analysis of issue types is limited by scale as manually annotating all 20,000 SWE-bench issues is infeasible, and many issues lack reliable labels. As a result, we do not claim that bug-fixing issues are inherently more prone to vulnerable patches; we only report that, within our manually annotated vulnerable instances, most affected issues are labeled as bugs.

Lastly, manual issue annotations may be prone to human error. To improve reliability, all annotations were independently performed by two reviewers, who ensured a strong inter-rater agreement (Cohen's Kappa $>$ 0.9).
\section{Discussion}
\label{sec:discussion}

\subsection{Standalone LLMs vs. Developers (RQ1)}

The RQ1 results show that, under our evaluated SWE-bench train+test setup, Llama 3.3 introduced substantially more retained new vulnerabilities than developers on the same issues, after majority voting and manual validation. The GPT-5.4 nano test-set results also show that manually confirmed vulnerabilities appear in a recent commercial standalone LLM, although the retained count is small and should be interpreted descriptively. Qualitatively, the retained vulnerabilities in Llama 3.3 train+test patches differ from those in the corresponding developer-written patches. Both include issues such as command injection and path traversal, but Llama 3.3 patches more frequently exhibit categories like eval injection and insecure deserialization, and in some cases downgrade existing cryptographic mechanisms or include multiple vulnerabilities within a single patch. GPT-5.4 nano also produced retained vulnerabilities in overlapping categories, including command injection, eval injection, and insecure deserialization. These patterns describe recurring security-relevant implementation choices in the retained Llama 3.3 patches under our evaluated APR setup.

SWE-bench Full contains instances with known quality issues that may affect the interpretation of our results. We cross-referenced our test-set results with the human annotations of SWE-bench Verified. Of the 2,294 test instances, 1,160 were marked for underspecified issue descriptions, problematic tests, or other issues affecting the functionality of the patches. These instances account for five of nine Llama vulnerabilities, ten of 44 OpenHands findings, three AutoCodeRover findings, and two developer findings in our test set. Although these benchmark-quality issues do
not necessarily invalidate the security vulnerabilities identified in the resulting patches, they may influence patch generation and limit the generalizability of our findings.

These findings indicate that standalone LLM-generated patches should be treated as an \textit{additional source of security risk.} From an adoption standpoint, the retained vulnerabilities in both evaluated standalone models support maintaining security checks for generated patches, including those proposed for mature projects. This implies that (i) generated patches should not bypass existing code review and security processes, (ii) organizations deploying LLM-based APR tools should consider integrating security scanning and some form of manual inspection into workflows that accept LLM patches, and (iii) the higher retained vulnerability count observed for Llama 3.3 motivates additional safeguards when considering its patches as replacements for developer-written repairs in the evaluated setting.

\subsection{Agentic APR Frameworks (RQ2)}

The RQ2 results show that agentic APR workflows built around LLMs do introduce security vulnerabilities, and that these vulnerabilities are not evenly distributed across frameworks or issues. In the evaluated setting, OpenHands produced a higher retained vulnerability count than AutoCodeRover, HoneyComb, and SWE-agent. In a small subset of OpenHands issues, we observed very large, multi-file diffs and multiple confirmed vulnerabilities, whereas the other evaluated frameworks generally modified fewer files and produced smaller patches. 

In the evaluated OpenHands setting, extensive edits accompanied a concentrated set of retained vulnerabilities. OpenHands produced 44 vulnerabilities across 11 issue instances, with 35 findings concentrated in three issues. These cases motivate further examination of how broad permissions to edit the repository, run arbitrary commands, and generate new scripts over hundreds of iterations relate to patch scope. The small number of affected issues limits generalization, and our study does not isolate autonomy, iteration control, or comprehension of the codebase as causes of the observed security outcomes.

The comparatively small number of vulnerabilities observed in AutoCodeRover, HoneyComb, and SWE-agent should not be interpreted as evidence that these systems are inherently secure or that tighter control over search space and modification scope caused their smaller retained counts. AutoCodeRover produced smaller patches than OpenHands in the evaluated data, motivating further study of edit scope as a potential risk indicator. At the same time, the vulnerabilities these frameworks introduce often fall into the same CWE categories, including command injection, eval injection, and insecure deserialization. The evaluated frameworks therefore share some vulnerability categories with the standalone models.

The results of RQ2 show that (i) retained OpenHands vulnerabilities are concentrated in a small number of issues involving extensive edits; (ii) AutoCodeRover, HoneyComb, and SWE-agent have smaller retained vulnerability counts in the evaluated setting; and (iii) the evaluated agentic frameworks share some vulnerability categories with developer-written patches, but the exact vulnerable locations differ. In our exact-overlap analysis, none of the agentic frameworks overlapped with developers on the same issue, same file, and same CWE. This emphasizes that these tools add a new layer of risk rather than merely repeating developers' mistakes.

For practitioners considering agentic APR tools, the inspected OpenHands cases motivate evaluating safeguards that monitor or limit edit scope, for example caps on diff size and limits on the number of files that can be modified in a single attempt. Their effectiveness in reducing vulnerabilities requires further evaluation. Additionally, given the concentration of retained OpenHands vulnerabilities in a small number of issues, practitioners could consider flagging repair tasks that repeatedly fail tests or produce unusually large diffs and routing them to human review.

\subsection{Code, Issue, and Project Characteristics of Vulnerable Instances (RQ3)}

Our analysis of over 20,000 LLM-generated patches and test-set agentic APR patches identifies associations between vulnerable outputs and observable characteristics, especially at the \textit{code} and \textit{issue} level.

At the \textit{code level}, we see a strong association between insecure patches and broader, more scattered edits. Vulnerable patches tend to touch more files and spread their changes across multiple locations, rather than making a self-contained change inside a single function or module. This pattern also holds descriptively for GPT-5.4 nano and SWE-agent. These are the situations where an LLM or APR agent must reason about cross-file dataflows and the interaction between different components, rather than a small local context. In the evaluated systems, vulnerable patches were associated with multi-file, dispersed edits. This association motivates evaluating whether constraining modifications to a single file or a specific component can improve patch security. When tasks require multi-file edits, it may be preferable to decompose the work into smaller changes, each operating over a part of the code, rather than requesting one large cross-file patch.

Other code-level factors show less consistent behavior. Llama and GPT-5.4 nano vulnerable patches generate fewer Python LOC on average than all generated patches, while SWE-agent show the opposite direction. Cyclomatic complexity is also mixed: AutoCodeRover vulnerable patches have substantially higher complexity, GPT-5.4 nano shows a descriptive increase, and SWE-agent shows a decrease. This suggests that file count is the most stable code-level signal in our analysis, while LOC and complexity are more framework-dependent. For some APR frameworks, complexity-sensitive safeguards, for example requiring human review or additional checks when edits occur in highly complex functions, may still be justified, even if the same signal is less informative elsewhere.

At the \textit{issue level}, vulnerabilities are commonly found in bugs where the issue descriptions lack information such as clear statements about the observed and expected behavior or steps to reproduce the bug. The newly annotated GPT-5.4 nano and SWE-agent instances mostly follow this pattern, but they also include security or deprecation-related feature requests. These observations motivate further study of issue information completeness as a contextual risk indicator. Across both Llama and the evaluated frameworks, higher word counts in issue descriptions do not strongly distinguish vulnerable from non-vulnerable instances. The contribution of information quality and relevance, including the usefulness of long stack traces or debugging conversations, warrants separate evaluation. At the same time, the GPT-5.4 nano and SWE-agent cases show that vulnerabilities can still occur even when issue reports include observed behavior, expected behavior, steps to reproduce, or code snippets. Vulnerabilities were observed even in cases with relatively complete issue reports.

At the \textit{project level}, we do not observe strong correlations between vulnerability introduction and repository size, code complexity, maintainability, or contributor count. This remains true after adding GPT-5.4 nano and SWE-agent. This may seem counterintuitive as one might expect that larger or more complex projects are inherently harder to repair securely. However, modern issue-tracking and modular design practices often constrain a given fix to a specific feature, module, or service, so many repairs remain local even in large systems. Conversely, a small codebase with tangled responsibilities and high cross-file coupling can force the model to reason across many interacting components even for a relatively simple problem.

RQ3 examines associations between vulnerable patch generation and repository, task, and patch characteristics, including repository size, the number of modified files, generated lines of code, and issue information completeness. We interpret the observed associations as contextual risk indicators, without treating them as causal explanations or evidence that a particular patch is vulnerable. These indicators are distinct from the semantic vulnerability patterns examined by static analysis, which concern unsafe operations and data-flow relationships, such as untrusted input reaching shell execution without appropriate sanitization. In our manually inspected cases, recurring security-relevant behaviors include replacing safe framework abstractions with manual implementations, using risky primitives such as \texttt{eval()}, and constructing commands, file paths, or SQL strings in ways that weaken existing safety checks. Contextual risk indicators may help prioritize patches for additional security checks or manual review, complementing the detection of specific vulnerabilities by static analysis.

Overall, the fact that vulnerability-prone patches correlate with a small set of observable code and task-related characteristics suggests a path toward lightweight, automated risk assessment for APR patches, analogous to prior work in abnormal commit detection~\cite{rosen2015commit, goyal2018identifying, leite2015uedashboard, gonzalez2021anomalicious}. Simple signals such as how widely a patch is spread across the codebase and the type of available information in an issue description can be combined into a prediction model that scores risk of APR usage \textit{before} and \textit{after} patch generation. In the input, issue-level signals can be used to flag high-risk tasks for APR systems. In the output, an additional set of code-level features can be used to evaluate generated patches. A risk model leveraging these factors would not replace static analysis or code review, but can be integrated into CI/CD pipelines as an early warning and triage mechanism where high-risk APR tasks or patches get routed to stricter security checks or human review. The signals needed for such risk assessment are already present in typical development workflows, making the integration of risk assessment into CI/CD a promising direction for future work.

\section{Related Work}
\label{related_work}
\indent

\textbf{Security Analyses of LLM-based APR Systems.}
Recent work has begun to analyze the security of LLM-based APR agents, rather than only their repair capabilities. Przymus et al. show that carefully crafted \textbf{adversarial bug reports} can mislead APR pipelines, inducing CVE reversions, vulnerability injection, CI/CD abuse, and denial-of-service even when some defenses such as AI-assisted reviews are deployed~\cite{przymus2025adversarialbugreportssecurity}. Their results highlight the bug-report interface as a new attack surface and suggest that current defenses block only a subset of malicious reports.

Chen et al. conduct a complementary red-teaming study of APR agents in realistic CI/CD settings, asking whether seemingly valid issue statements can cause agents to generate vulnerable patches~\cite{chen2025redteamingprogramrepair}. They challenge the common assumption that tests passing implies safety for APR-generated patches, and demonstrate that agents can be compelled into introducing exploitable behavior while still satisfying existing test suites.

Both works build on a rapidly growing ecosystem of LLM-based APR agents which primarily focus on improving repair success and benchmark performance in repository-scale settings, often without any security checks~\cite{ruiz2025art, yang2025enhancing, zhao2024enhancing, fan2023automated, zhang2024autocoderover, openhands, honeycomb}. In contrast to these adversarial perspectives, our study analyzes LLM and agentic APR outputs under \textbf{non-adversarial} SWE-bench issues, quantifying how often vulnerabilities arise by accident and linking them to code and task-level factors such as edit scope and issue information completeness, as well as comparing them directly to developer-written patches.

\textbf{Security Analyses of LLM-generated Code.} Given the growing adoption of LLMs by developers \cite{daigle2024aiwave, shani2023aiimpact, kabir2024stack}, recent studies have begun to examine various aspects of LLM-generated code \cite{fan2023large, liu2024exploring, jimenez2024swebench, wang2025solved}. While many of these studies have focused on functionality, others have examined the security implications of LLM-generated code \cite{pearce2025asleep, fu2023security, toth2024llms, khoury2023secure, siddiq2024quality, mohsin2024can}.
Pearce et al. \cite{pearce2025asleep} conducted one of the first assessments on the security of LLM-generated code, showing that even state-of-the-art models can produce vulnerable code in security-sensitive scenarios. Since then, many studies have confirmed these findings by analyzing LLM-generated code across different tasks, models, and user interaction settings \cite{ramirez2024state, perry2023users, liu2024refining, siddiq2023generate}. 

Most prior work has focused on small code snippets or controlled experimental settings \cite{mohsin2024can, siddiq2024quality, pearce2025asleep, khoury2023secure}. These settings often lack the full project context and complexity inherent in real-world software maintenance tasks, limiting their ability to assess how vulnerabilities emerge in practice. However, a recent study by Fu et al.~\cite{fu2023security} analyzed Copilot-generated code snippets in GitHub projects and identified several security weaknesses. It is worth mentioning that because these completions are reviewed and integrated by developers, the model is not fully in control of the final output. Our work complements and extends this work by analyzing LLM-generated code in real-world GitHub issues, where models are given autonomy to modify any relevant files. To our knowledge, this is the first large-scale empirical study to link vulnerabilities introduced by LLMs and agentic frameworks to a broad set of contextual factors, including issue descriptions, code-level attributes, and project structure, in realistic software development settings where models are tasked with resolving issues.  

\textbf{Vulnerability Detection.} Automated vulnerability detection can be performed using static or dynamic analysis, with static analysis being widely adopted for large-scale, code-centric studies. Static analysis detects weaknesses by reasoning about source code without execution through techniques such as rule/pattern matching, taint and dataflow analysis, and abstract interpretation~\cite{owasp_source_code_analysis, sajadi2026empiricalanalysiscodeqlfalse}. Tools like CodeQL, Semgrep, and SpotBugs leverage these techniques with CWE-based queries and rules that support automated evaluations of code~\cite{GitHubCodeQL, semgrep, spotbugs}. More recent work has explored leveraging LLMs for detection, either directly, by identifying vulnerabilities in code snippets or projects~\cite{ullah2024llms, chakraborty2021deep}, or indirectly by enhancing static analysis pipelines~\cite{sheng2025llms, lu2024grace}. In the latter case, LLMs have been used for tasks such as inferring vulnerability-specific taint specifications, identifying function call chains, and reasoning over dataflow paths to improve static taint analysis~\cite{li2025iris, li2025hitchhiker, liu2025llm}. While LLM-based detection introduces new opportunities, established static analysis tools remain widely used in practice, particularly in CI/CD pipelines, where their deterministic query semantics and reproducibility are often essential.

\textbf{Abnormal Commit Detection.} Past studies have explored various techniques to identify and characterize the risk of potential changes in software systems, particularly focusing on GitHub commits \cite{rosen2015commit, goyal2018identifying, leite2015uedashboard, gonzalez2021anomalicious}. This line of research will generally aim to identify outlier commits, also known as risky or abnormal commits, through assessing various commit attributes. For instance, Gonzalez et al. introduced Anomalicious \cite{gonzalez2021anomalicious}, an approach to identify malicious commits. By analyzing the commits and identifying the characteristics of a ``normal" developer commit, they made an attempt at detecting malicious code contributions.

While these studies provide valuable foundations for identifying risky, malicious, or insecure commits, they have been developed to evaluate human-written code. Our work is the first to assess the characteristics of insecure code produced by LLMs and agentic frameworks. By comparing vulnerable and non-vulnerable generated patches across a range of factors (e.g., scope of modifications, issue type, information completeness) we identify the unique characteristics of high-risk agentic or LLM-generated commits.

\section{Conclusions}

This study presents a large-scale security evaluation of standalone LLM-generated and agentic APR-generated patches for real-world issue resolution tasks. Under our SWE-bench train+test evaluation setup, Llama 3.3 introduced 135 manually confirmed new vulnerabilities, approximately 11 times the 12 introduced by developers on the same issues. We also find that GPT-5.4 nano, evaluated on the SWE-bench test set, introduces manually confirmed vulnerabilities, so vulnerability introduction was observed in both evaluated standalone models. Llama 3.3 produced more retained findings in categories such as CWE-95 and CWE-502 than developers in this setup; the corresponding developer patches also included categories such as CWE-732 and CWE-377. Further, our analysis of agentic frameworks shows that all four evaluated frameworks introduce vulnerabilities, but the risk is unevenly distributed. OpenHands produces the largest number of retained vulnerabilities, while AutoCodeRover, HoneyComb, and SWE-agent produce fewer retained vulnerabilities. OpenHands produced 44 retained vulnerabilities across 11 issue instances, with 35 findings concentrated in three issues. These observations are specific to the evaluated setting and do not establish that increasing autonomy caused the differences between frameworks.

Our findings identify correlations between vulnerable outputs and particular code and issue-level characteristics. The most consistent code-level signal is the number of modified files, while unique file types, generated LOC, and cyclomatic complexity show more mixed patterns across systems. At the issue level, missing code snippets, missing expected behavior, and missing reproduction steps were observed among vulnerable instances, while the GPT-5.4 nano and SWE-agent cases show that vulnerabilities can still occur even when issue reports are relatively complete. These patterns suggest that the riskiness of LLM or agent-generated patches can be assessed not only by detecting specific vulnerable lines of code, but also by developing complementary risk assessment techniques that evaluate both the generated code and the broader context in which a patch is proposed, such as issue type, information completeness, and edit scope. Such systems can complement existing vulnerability detection techniques, often reliant on costly manual reviews, by narrowing the focus to potentially high-risk code contributions.

\appendix
\section*{Appendix}
\section{Full CWE Distribution for RQ1}
\label{app:rq1-full-cwe}

\begin{longtable}{p{0.24\linewidth}p{0.12\linewidth}p{0.48\linewidth}r}
\caption{Full CWE distribution for manually retained RQ1 vulnerabilities. Counts are over retained vulnerability rows. Each row can contribute multiple CWE IDs when tools report multiple CWEs.}
\label{tab:rq1_full_cwe_distribution} \\
\toprule
\textbf{Source} & \textbf{CWE} & \textbf{Short Name} & \textbf{Count} \\
\midrule
\endfirsthead

\toprule
\textbf{Source} & \textbf{CWE} & \textbf{Short Name} & \textbf{Count} \\
\midrule
\endhead

\bottomrule
\endfoot

\textbf{Llama 3.3, Train+Test} & CWE-78 & OS Command Injection & 79 \\
 & CWE-95 & Eval Injection & 39 \\
 & CWE-502 & Insecure Deserialization & 17 \\
 & CWE-22 & Path Traversal & 14 \\
 & CWE-939 & Improper Authorization in Handler for Custom URL Scheme & 8 \\
 & CWE-732 & Permission Misconfiguration & 6 \\
 & CWE-276 & Incorrect Default Permissions & 5 \\
 & CWE-94 & Code Injection & 4 \\
 & CWE-215 & Sensitive Information in Debug Code & 4 \\
 & CWE-259 & Hard-coded Password & 4 \\
 & CWE-489 & Active Debug Code & 4 \\
 & CWE-209 & Error Message Containing Sensitive Information & 3 \\
 & CWE-312 & Cleartext Storage of Sensitive Information & 3 \\
 & CWE-321 & Hard-coded Cryptographic Key & 3 \\
 & CWE-327 & Broken or Risky Crypto Algorithm & 3 \\
 & CWE-359 & Exposure of Private Personal Information & 3 \\
 & CWE-497 & Exposure of Sensitive System Information & 3 \\
 & CWE-532 & Sensitive Information in Log File & 3 \\
 & CWE-798 & Hard-coded Credentials & 3 \\
 & CWE-20 & Improper Input Validation & 2 \\
 & CWE-79 & Cross-site Scripting & 2 \\
 & CWE-328 & Weak Hash & 2 \\
 & CWE-377 & Insecure Temporary File & 2 \\
 & CWE-703 & Improper Exception Handling & 2 \\
 & CWE-706 & Incorrectly Resolved Name or Reference & 2 \\
 & CWE-776 & XML Entity Expansion & 2 \\
 & CWE-916 & Insufficient Password Hashing Effort & 2 \\
 & CWE-23 & Relative Path Traversal & 1 \\
 & CWE-36 & Absolute Path Traversal & 1 \\
 & CWE-73 & External Control of File Name or Path & 1 \\
 & CWE-89 & SQL Injection & 1 \\
 & CWE-99 & Resource Injection & 1 \\
 & CWE-295 & Improper Certificate Validation & 1 \\
 & CWE-319 & Cleartext Transmission of Sensitive Information & 1 \\
 & CWE-400 & Uncontrolled Resource Consumption & 1 \\
 & CWE-601 & Open Redirect & 1 \\
\midrule

\textbf{Developers, Train+Test} & CWE-78 & OS Command Injection & 4 \\
 & CWE-22 & Path Traversal & 3 \\
 & CWE-377 & Insecure Temporary File & 3 \\
 & CWE-732 & Permission Misconfiguration & 3 \\
 & CWE-502 & Insecure Deserialization & 2 \\
 & CWE-79 & Cross-site Scripting & 1 \\
 & CWE-94 & Code Injection & 1 \\
 & CWE-95 & Eval Injection & 1 \\
\midrule

\textbf{GPT-5.4 nano, Test} & CWE-78 & OS Command Injection & 1 \\
 & CWE-95 & Eval Injection & 1 \\
 & CWE-502 & Insecure Deserialization & 1 \\
 & CWE-703 & Improper Exception Handling & 1 \\
 & CWE-706 & Incorrectly Resolved Name or Reference & 1 \\
\midrule

\textbf{Developers, Test} & CWE-78 & OS Command Injection & 1 \\
 & CWE-95 & Eval Injection & 1 \\
 & CWE-502 & Insecure Deserialization & 1 \\

\end{longtable}

\begin{acks}
We acknowledge the support of the National Science Foundation under grants 2622910 and 2622911.
\end{acks}


\end{document}